\documentclass{article}

\usepackage{cite}
\usepackage{amsmath,amssymb,amsfonts}
\usepackage{graphicx}
\usepackage{textcomp}
\def\BibTeX{{\rm B\kern-.05em{\sc i\kern-.025em b}\kern-.08em
    T\kern-.1667em\lower.7ex\hbox{E}\kern-.125emX}}

\usepackage{algorithm}
\usepackage{textcomp}
\usepackage{stfloats}
\usepackage{url}
\usepackage{verbatim}
\usepackage{soul}
\usepackage{color}
\usepackage{hyperref}
\usepackage{amsmath}
\usepackage{xcolor}
\usepackage{xspace}
\usepackage{mathtools}
\usepackage{algorithmicx}
\usepackage[noend]{algpseudocode}

\usepackage{PRIMEarxiv}

\usepackage[utf8]{inputenc} 
\usepackage[T1]{fontenc}    
\usepackage{hyperref}       
\usepackage{url}            
\usepackage{booktabs}       
\usepackage{amsfonts}       
\usepackage{nicefrac}       
\usepackage{microtype}      
\usepackage{lipsum}
\usepackage{fancyhdr}       
\usepackage{graphicx}       
\graphicspath{{media/}}     

\newcommand\mn{{DeepTextMark}\xspace}

\title{DeepTextMark: A Deep Learning-Driven Text Watermarking Approach for Identifying Large Language Model Generated Text
}

\author{
  Travis Munyer, Abdullah Tanvir, Arjon Das, Xin Zhong \\
  Department of Computer Science \\
  University of Nebraska Omaha \\
  Omaha, NE 68182 USA \\
  \texttt{tjmunyer@gmail.com, \{atanvir, arjondas, xzhong\}@unomaha.edu} \\
}

\begin{document}
\maketitle

\begin{abstract}
The rapid advancement of Large Language Models (LLMs) has significantly enhanced the capabilities of text generators. 
With the potential for misuse escalating, the importance of discerning whether texts are human-authored or generated by LLMs has become paramount. 
Several preceding studies have ventured to address this challenge by employing binary classifiers to differentiate between human-written and LLM-generated text. Nevertheless, the reliability of these classifiers has been subject to question.
Given that consequential decisions may hinge on the outcome of such classification, it is imperative that text source detection is of high caliber.
In light of this, the present paper introduces \mn, a deep learning-driven text watermarking methodology devised for text source identification.  
By leveraging Word2Vec and Sentence Encoding for watermark insertion, alongside a transformer-based classifier for watermark detection, \mn epitomizes a blend of blindness, robustness, imperceptibility, and reliability.
As elaborated within the paper, these attributes are crucial for universal text source detection, with a particular emphasis in this paper on text produced by LLMs.
\mn offers a viable "add-on" solution to prevailing text generation frameworks, requiring no direct access or alterations to the underlying text generation mechanism.
Experimental evaluations underscore the high imperceptibility, elevated detection accuracy, augmented robustness, reliability, and swift execution of \mn.

\end{abstract}


\section{Introduction}
\label{intro}
Large Language Models (LLMs), such as ChatGPT~\cite{chatgpt}, have recently achieved notable success. 
The advancements in LLMs can be advantageous across various domains, yet there also lies the potential for inappropriate applications. 
A prevailing concern regarding publicly accessible LLMs is the challenge in distinguishing between machine-generated and human-written text, a difficulty that persists even in instances of misuse~\cite{OpenAIBlog}. 
For instance, students might utilize automatically generated texts as their own submissions for assignments, evading conventional "plagiarism" detection. 
The high fidelity of the text generated by LLMs exacerbates the challenge of detection, marking a significant hurdle. 
Again, there exist advanced text augmentation methods capable of effortlessly modifying any given text \cite{onan2023gtr}\cite{onan2024improving}\cite{onan2023srl}\cite{onan2022bidirectional}.
Therefore, devising a method to ascertain the origin of text could serve as a valuable approach to curtail similar misapplications of LLMs.

\begin{figure}[t!]
    \centering
    \includegraphics[width=0.5\linewidth]{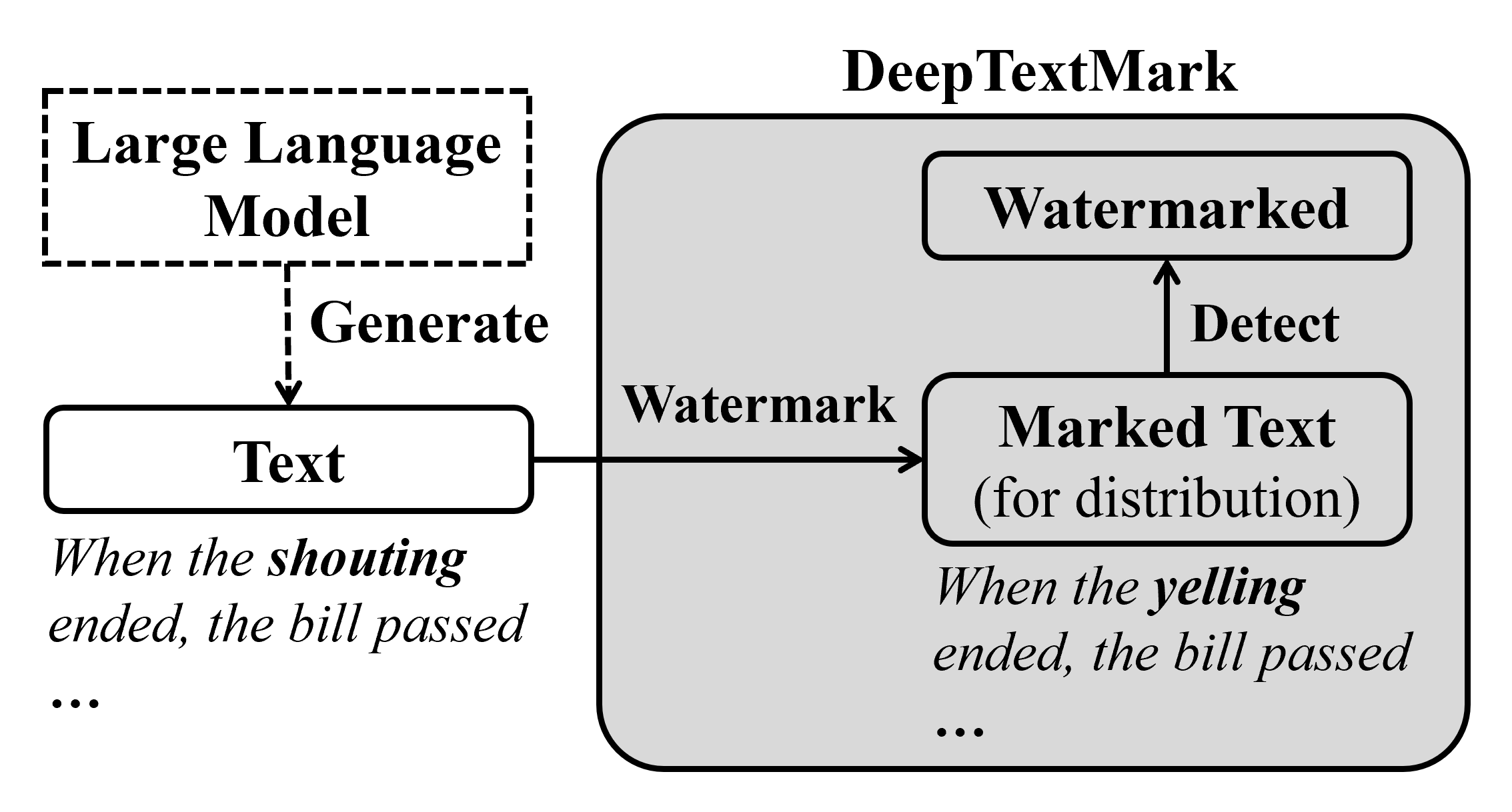}
    \vspace{-0.3em}
    \caption{Overall idea of \mn.}
    \vspace{-1.7em}
    \label{fig:Overall}
\end{figure}

Various classifiers have been developed to differentiate between LLM-generated text and human-written text~\cite{OpenAIBlog, gptzero}. 
However, the efficacy of these classifiers remains somewhat constrained at present. 
Numerous studies have explored the accuracy of these classifiers~\cite{ajaay_2023}, along with techniques to circumvent classifier detection~\cite{sheinman_2023}. 
A reliable source detection mechanism that is challenging to bypass is crucial, given its potential applications in identifying plagiarism and misuse. 
Therefore, employing text watermarking for text source detection appears to be a prudent approach, as it is both reliable and challenging to circumvent.

Text watermarking entails the covert embedding of information (\textit{i.e.}, the watermark) into cover texts, such that the watermark is only discernible by authorized detectors. While watermarking is more conventionally applied to images~\cite{fang2022pimog}, its application to text can enable the identification of text originating from specific sources, such as an LLM (refer to figure~\ref{fig:Overall} for the proposed source detection mechanism).
However, conventional text watermarking techniques often necessitate manual intervention by linguists, exhibit a lack of robustness, and do not possess the blindness property.
Specifically, these traditional techniques are prone to minor modifications of the watermarked text (lacking robustness), and necessitate the original text for the extraction or detection of the watermark (lacking blindness).
For a watermarking technique to be practically viable in detecting LLM-generated text, the method should be scalable (\textit{i.e.}, automatic).
Moreover, since the watermark detector may not have access to the original text at the time of detection, it should not require it (\textit{i.e}., blind).
Additionally, the detection process should be highly reliable, aiming to achieve superior classification accuracy.
Ideally, the watermarked text should remain imperceptible, ensuring the natural preservation of the text's meaning.
Lastly, the classification mechanism should be resilient to minor alterations of the text (\textit{i.e.}, robust).

A nascent method has been proposed for embedding watermarks into LLMs~\cite{kirchenbauer2023watermark}. However, a notable limitation of this method is its requisite access to the text generation phase of the LLMs, a requirement that may not be practical in real-world applications, particularly when the source models of the LLMs are not open-source. This dependency poses a significant challenge as many LLMs are proprietary or their internals are not publicly disclosed, thereby restricting the applicability of such watermarking techniques. Moreover, without the requisite access to the text generation phase, implementing watermark-based source detection mechanisms becomes inherently challenging. This highlights the necessity for developing alternative watermarking techniques that are both effective and adaptable to varying levels of access to the LLMs' internal workings.

This paper introduces \mn, a robust and blind deep learning-based text watermarking method principally aimed at detecting LLM-generated text. 
\mn employs word substitution, utilizing a pre-trained amalgam of Universal Sentence Encoder embeddings~\cite{cer2018universal} and Word2Vec~\cite{mikolov2013distributed} to identify semantically congruent substitution words, thereby maintaining the imperceptibility of the watermarked text. 
Moreover, we propose a novel classifier, grounded in transformer architecture~\cite{10.5555/3295222.3295349}, to discern watermarked text, enhancing detection accuracy and robustness. 
This amalgam of pre-trained models for substitution word selection and the transformer-based watermark detector underscore the novel contributions of this paper. 
Being deep learning-driven, the watermarking and detection techniques are scalable and fully automatic. 
The classifier necessitates only the watermarked text for highly accurate classification, epitomizing the technique's blindness. 
Furthermore, the paper elucidates an extension of this technique to multiple sentences, like essays, accentuating a primary application. Empirical evidence is provided demonstrating near-perfect accuracy as text length increases, enriching the method's reliability, especially with a modest sentence count. 

The primary contributions encapsulate: (1) an "add-on" text watermarking method for detecting generated text without necessitating access to the LLMs' generation phase; (2) an automatic and imperceptible watermark insertion method; and (3) a robust, high-accuracy deep learning-based text watermark detection method, rendering \mn a valuable asset in the realm of text authenticity verification.

The rest of this paper is organized as follows. We discuss related works in section~\ref{related}. The watermark insertion and detection process is discussed in section~\ref{method}. Experiments showing the reliability, imperceptibility, robustness, and empirical runtime are shown in section~\ref{experiments} followed by a conclusion of the work in section~\ref{conclusion}.

\section{Related Work}
\label{related}
Our contributions are summarized as robust detection of LLM-generated text, a novel method for text watermarking insertion, and a novel approach for text watermarking detection; the following sections provide a review of related work in these domains. 
Section~\ref{subsec:RelatedLLM} offers a concise review of state-of-the-art methods for LLM-generated text detection, while Section~\ref{subsec:ClassicalTextWm} delves into classical text watermarking techniques.

\subsection{Text Source Detection for Large Language Models}
\label{subsec:RelatedLLM}
Recent endeavors have been directed towards developing classifiers aimed at differentiating between LLM-generated text and human-written text. 
The prevailing approach entails the collection and labeling of LLM-generated and human-written texts, followed by the training of a binary classifier through supervised learning. 
Although the efficacy of these classifiers has yet to be fully established, some preliminary analyses have been reported~\cite{sheinman_2023, ajaay_2023}. 
One study~\cite{sheinman_2023} elucidated three distinct methods, substantiated with examples, to circumvent the GPTZero~\cite{gptzero} classifier detection. 
Another investigation~\cite{ajaay_2023} conducted a direct assessment of GPTZero's accuracy, uncovering inconsistencies in its ability to detect human-written text. 
Moreover, classifier-based LLM-generated text detectors commonly necessitate a substantial character count to perform detection accurately. 
For instance, GPTZero~\cite{gptzero} required a minimum of 250 characters to initiate detection. 
Looking ahead, OpenAI is planning a cryptography-based watermarking system for ChatGPT-generated text detection~\cite{montti_2022}, although no definitive work has been disclosed as of yet. 
Zero-shot learning-based methods have also demonstrated some advancement. 
For example, Cer \textit{et al}.~\cite{mitchell2023detectgpt} reported an increment in AUROC from $1$\% to $14$\% compared to other zero-shot detection strategies across various datasets; however, the accuracy might still fall short in real-world applications concerning text generated by models.

A method has been proposed for detecting LLM-generated texts based on text watermarking~\cite{kirchenbauer2023watermark}, which involves watermarking the text by modifying the LLMs (sensitive tokens are defined and excluded from the output of the LLMs).
In contrast, our proposed \mn does not necessitate access to or modifications of the LLM.
Distinct from model-dependent methods, \mn exhibits a model-independent feature, enabling its application to any text.
Moreover, \mn employs a substantially more compact architecture with about 50 million parameters, whereas the method in~\cite{kirchenbauer2023watermark} necessitates billions of parameters to implement the watermarking process.

A pertinent topic in text watermarking for identifying generated text is the potential use of paraphrasing attacks to bypass AI-detectors, as elaborated in a study by Sadasivan \textit{et al.}~\cite{sadasivan2023aigenerated}.
This concern is not relevant to our target scenario, as \mn focuses solely on the detection of text output by an LLM.
Should a human writer meticulously rewrite the text generated by an LLM, the resultant paraphrased text may not be subject to "plagiarism" detection in our scenario.

Relative to existing state-of-the-art methods, our proposal exhibits several advantages:
(1) Our watermarking method renders detection bypass challenging unless the LLM-generated text is rewritten, as the watermark is embedded in undisclosed locations, necessitating a rewrite for its removal. Once rewritten, the text is deemed as distinct human-written text;
(2) The method demonstrates high detection accuracy, nearing 100\%, which significantly elevates with an increasing number of sentences, substantiated through binomial distribution analysis. Even on a single sentence, a reliable detection rate of 86.52\% is achieved;
(3) To our knowledge, this is the inaugural LLM-independent, deep learning-based general text watermarking method;
(4) Unlike some methods necessitating access to text generation processes, our approach requires no access to the LLM's original text generation, allowing our watermarker to function as an "add-on" to the LLM system (see Figure~\ref{fig:Overall}).

\subsection{Traditional Text Watermarking}
\label{subsec:ClassicalTextWm}
Common classical text watermarking methods can be categorized into open space, syntactic, semantic, linguistic, and structural techniques. A brief summary of each of these techniques is provided below.

\noindent \textbf{Open Space:} 
The open space method embeds a watermark into text data by adding extra white space characters or spaces at specific locations in the text~\cite{ou_2003}. 
For instance, extra white space between words or lines could be encoded as a 1, while normal white space could encode as a 0. 
The strategy for adding extra white space and its encoding is subject to the implementation.
Although the open space method can be simple to implement and automate, it may be susceptible to watermark removal without altering the text's meaning, as an individual could easily eliminate the extra white space.

\noindent \textbf{Syntactic:} 
Certain word orders can be altered without changing the meaning or grammatical correctness of a sentence. The syntactic method watermarks text by modifying the order of words in sentences~\cite{TextWatermarkingReview}. 
For example, "this and that" could encode to 1, and "that and this" could encode to 0. However, this method may not scale well since many sentences do not have sequences of words that qualify for reordering. 
Additionally, this method might necessitate manual intervention by a linguist, as developing an automated system to detect reorderable words could be challenging.

\noindent \textbf{Semantic:} 
Semantic text watermarking techniques embed the watermark by substituting words with synonyms~\cite{TextWatermarkingReview}. 
While the semantic method can be automated, as briefly discussed in this paper, classical implementation requires the original text to detect the watermark (i.e., classical semantic text watermarking is non-blind). 
Moreover, determining which word to replace, and selecting an appropriate synonym, presents a non-trivial challenge.

\noindent \textbf{Linguistic:} 
The linguistic category of text watermarking amalgamates semantic and syntactic techniques, embedding watermarks into text through a blend of word rearrangement and synonym replacement~\cite{TextWatermarkingReview}.

\noindent \textbf{Structural:} 
The structural technique replaces certain symbols with visually similar letters and punctuation, albeit with different Unicode representations~\cite{Rizzo2019}. 
It may be relatively straightforward to detect these symbols either manually due to minor visual differences, or automatically by identifying characters from uncommon Unicode sets. 
Reverting the watermarking without altering the text's meaning could also be straightforward. 
Due to these limitations, structural techniques do not align with our primary objective of watermarking text generated by LLMs.

Contrastingly, we employ word2vec~\cite{mikolov2013distributed} and the Universal Sentence Encoder~\cite{cer2018universal} for watermark insertion, and devise a transformer-based model for watermark detection. 
This approach aligns well with our target application of text source detection, as it facilitates blindness while enhancing imperceptibility, robustness, and reliability. 
Our watermark insertion and detection methodology is rooted in deep learning, distinguishing our method from traditional text watermarking techniques.

\section{The Proposed \mn}
\label{method}
This section presents the details of \mn. 
The proposed watermark insertion and detection schemes are respectively discussed in Sections~\ref{subsec:EmbeddingMethod} and~\ref{subsec:ExtractionMethod}. 
This discussion shows the automatic and blindness traits achieved by \mn. Section~\ref{subsec:ProposedSentenceCollection} analyzes the application scenario of \mn to multiple sentences. 

\subsection{Watermark Insertion}
\label{subsec:EmbeddingMethod}

\begin{figure*}[ht!]
    \centering
    \includegraphics[width=0.85\linewidth]{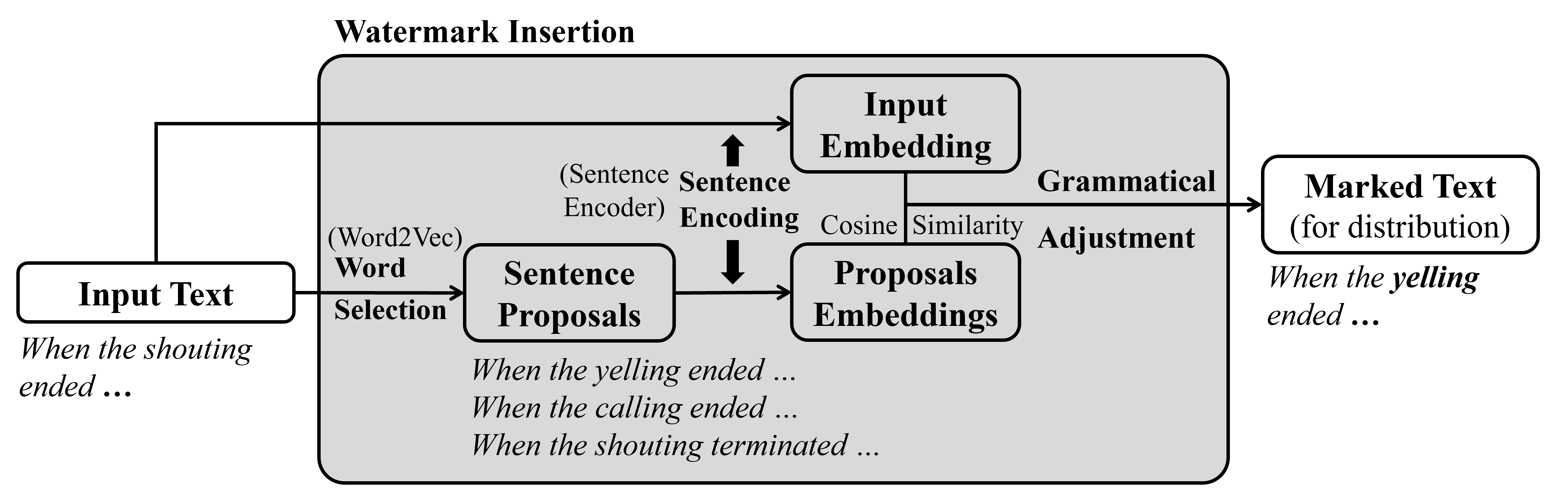}
    \vspace{-0.6em}
    \caption{Watermark insertion details.} 
    \vspace{-0.6em}
    
    \label{fig:insertion}
\end{figure*}

In contemporary settings, individuals employ extensive language models to produce textual content and subsequently rephrase it using synonymous words as a strategy to circumvent plagiarism. This serves as the rationale behind our introduced watermark insertion model, aiming to detect alterations in text even when someone attempts to paraphrase content generated by large language models in order to evade plagiarism detection. The watermark insertion process is presented in Figure~\ref{fig:insertion}. 

\noindent \textbf{Word Selection:} 
Given a sentence, we initially segregate candidate words from punctuation, stopwords~\cite{7375527}, and whitespace, preserving these elements to retain the original sentence structure. Each candidate word is then transposed to an embedding vector utilizing a pre-trained Word2Vec model~\cite{mikolov2013distributed}.
A roster of replacement words is engendered by identifying the $n$ nearest vectors to the candidate word vector in the Word2Vec embedding space, where $n$ is a pre-defined integer, and reconverting these vectors back into words.
We engender a list of sentence proposals by substituting each candidate word with its list of replacement words, thereby fabricating unique sentence variations.
The loci of the watermark in each sentence proposal are indirectly ascertained by Word2Vec.
Each unique variation is deemed a sentence proposal, representing a potential watermarked sentence.
Empirically, employing a larger corpus of nearest vectors allows for the consideration of an augmented set of replacement words and consequently more sentence proposals, potentially ameliorating imperceptibility albeit at the expense of elevated processing time.
We also delved into various word-level watermarking techniques. Initially, a sole word within each sentence was substituted with its synonyms which we denote as \textbf{single word synonym substitution}. This scope was subsequently broadened to encompass multiple-word replacements within each sentence which is denoted by \textbf{multiple word synonyms substitution}. In the terminal phase of our experimentation, we embraced a flexible approach, permitting the substitution of any candidate word with an available synonym in a sentence.

\noindent \textbf{Sentence Encoding:} 
At this juncture, each sentence proposal is evaluated solely based on word-level quality. 
We ascertain that the quality of the watermarked sentence is enhanced when the architecture is allowed to consider sentence-level quality. 
To facilitate this, we employ a pretrained Universal Sentence Encoder~\cite{cer2018universal} to score the quality of each sentence proposal. 
This encoder transposes a sentence into a high-dimensional vector representation. 
Initially, both the original sentence and each sentence proposal are transposed into vector representations using the Universal Sentence Encoder. 
Subsequently, we compute the similarity score for each sentence proposal by measuring the cosine similarity between the vector representation of the original sentence and that of the sentence proposal. 
The sentence proposal exhibiting the highest similarity score is identified as the potential watermarked sentence. 
Given that the watermarking process necessitates no human intervention, the methodology is rendered automatic.

\noindent \textbf{Grammatical Adjustment:} 
\label{subsec:GrammaticalError}
In pursuit of mitigating grammatical inaccuracies, essential measures have been undertaken. Our methodology encompasses word substitution with synonymous counterparts, whilst steadfastly preserving the original sentence structure. In this vein, we have eschewed the elimination of stopwords or the alteration of punctuation, thereby safeguarding sentence integrity.

The process of synonym selection is meticulously designed to favor optimal replacements. Nevertheless, challenges emerge in instances where the most apt synonym diverges in grammatical structure or meaning. For instance, replacing the term 'elections,' a plural noun, with 'election,' its singular counterpart, could engender grammatical incongruity. 
To forestall such scenarios, a preliminary determination of the grammatical number of the target word is initiated with a class engine~\cite{inflect} which employs diverse methods to facilitate plural and singular inflections, the selection of "a" or "an" for English words based on pronunciation, and the manipulation of numbers represented as words. This module comprehensively provides plural forms for nouns, most verbs, and select adjectives, including "classical" variants like transforming "brother" to "brethren" or "dogma" to "dogmata." Singular forms of nouns are also available, allowing the choice of gender for singular pronouns, such as transforming "they" to "it," "she," "he," or "they." Pronunciation-based "a" or "an" selection is extended to all English words and most initialisms. It is crucial to note that when using plural inflection methods, the word to be inflected should be the first argument, expecting the singular form; passing a plural form may yield undefined and likely incorrect results. Similarly, the singular inflection method anticipates the plural form of the word. The plural inflection methods also offer an optional second argument indicating the grammatical "number" of the word or another word for agreement. Subsequently, synonyms congruent with the grammatical form of the original word are curated. 

\begin{table}[H]
\caption{Example Sentence Candidates of Correct and Incorrect Synonym Selections}
\vspace{-0.5em}
\centering
\begin{tabular}{|p{7.5cm}|} 
\hline
1. The September-October \textbf{term} jury had been charged by Fulton Superior Court judge Durwood Pye to investigate reports of possible "irregularities" in the hard-fought primary which was won by mayor-nominate Ivan Allen Jr. \\
\hline
2. The September-October \textbf{terms} jury had been charged by Fulton Superior Court judge Durwood Pye to investigate reports of possible "irregularities" in the hard-fought primary which was won by mayor-nominate Ivan Allen Jr. \\
\hline
3. The September-October \textbf{condition} jury had been charged by Fulton Superior Court judge Durwood Pye to investigate reports of possible "irregularities" in the hard-fought primary which was won by mayor-nominate Ivan Allen Jr. \\
\hline
\end{tabular}
\label{tbl:CandidateSentences}
\end{table}

A few examples of sentence candidates with correct and incorrect synonym selections are presented in Table~\ref{tbl:CandidateSentences}. 
It is imperative to note that when we scrutinize the word \textbf{term}, we encounter the closest synonyms, some of which contravene the grammatical criteria due to their distinct grammatical numbers, with one being singular and the other plural. Consequently, given that our initial word is in the singular form, our consideration is limited exclusively to synonyms in the singular form. Consequently, in lieu of employing \textbf{terms}, we opt to substitute it with \textbf{condition}. 

Analogous complexities arise concerning parts of speech, as certain words harbor synonyms across diverse lexical categories. To adeptly navigate this intricacy, integration of the classic POS (Part of Speech) tagger~\cite{nltk} has been effected. Post identification of the word's grammatical number, the endeavor to pinpoint synonyms aligning with its specific part of speech is undertaken. This bifurcated approach underpins both syntactic and grammatical consistency in our synonym substitution process. 

\begin{table}[H]
\caption{Example Sentence Candidates with Varied Synonyms}
\vspace{-0.5em}
\centering
\begin{tabular}{|p{7.5cm}|} 
\hline
The \textbf{primary} thing she did was to take off her hat and then as she had no other covering she. \\
\hline
The \textbf{first} thing she did was to take off her hat and then as she had no other covering she. \\
\hline
The \textbf{leading} thing she did was to take off her hat and then as she had no other covering she. \\
\hline
\end{tabular}
\label{tbl:VariedSynonymCandidates}
\end{table}

A few examples of sentence candidates with varied synonyms selections are presented in Table~\ref{tbl:VariedSynonymCandidates}. 
An analysis of the term \textbf{primary} reveals that the closest synonyms are typically adverbs like \textbf{first}, thereby deviating from the grammatical condition, as the original term is a proper noun (NNP). Given the categorical distinction that our original word falls into the proper noun category (NNP), our focus is exclusively on synonyms that share this grammatical property. This rationale informs our decision to replace the word \textbf{primary} with \textbf{leading} instead of \textbf{first}. 
We have implemented the type of parts of speech by using the POS-tagger provided by the NLTK~\cite{nltk}. Specifically, we employed the Penn Treebank POS tagger. The tagging process involved tokenization of input text, breaking it into individual words or sentences, and subsequently assigning part-of-speech tags to each word. The POS tagging was conducted using a Hidden Markov Model (HMM), trained on a large annotated corpus, such as the Penn Treebank corpus, wherein the model learned the probabilities of transitions between different POS tags and the probabilities of observing specific words given a certain POS tag. The Viterbi algorithm was employed during the tagging of new text to identify the most likely sequence of POS tags given the observed words and the learned probabilities. This approach proved effective for obtaining accurate and contextually relevant part-of-speech annotations in diverse textual datasets. Algorithm~\ref{algo:insertion_algo} outlines the entire operational process.

\begin{algorithm}
\caption{Watermark Insertion}
\label{algo:insertion_algo}
\begin{algorithmic}[1]
\Function{WatermarkInsertion}{input\_text}
\State word\_embedder $\gets$ \Call{Word2Vec}{}
\State sentence\_encoder $\gets$ \Call{SentenceEncoder}{}
\State input\_embeddings $\gets$ \Call{Encode}{word\_embedder, input\_text}
\State sentence\_proposals $\gets$ \Call{GenerateProposals}{input\_text}
\State proposals\_embeddings $\gets$ \Call{Encode}{sentence\_encoder, sentence\_proposals}
\State best\_proposal $\gets$ \Call{ComputeCosineSimilarity}{input\_embeddings, proposals\_embeddings}
\State marked\_text $\gets$ \Call{GrammaticalAdjustment}{best\_proposal}
\State \Return marked\_text
\EndFunction
\end{algorithmic}
\end{algorithm}




\subsection{Watermark Detection}
\label{subsec:ExtractionMethod}

\begin{figure*}[h!]
    \centering
    \includegraphics[width=0.65\linewidth]{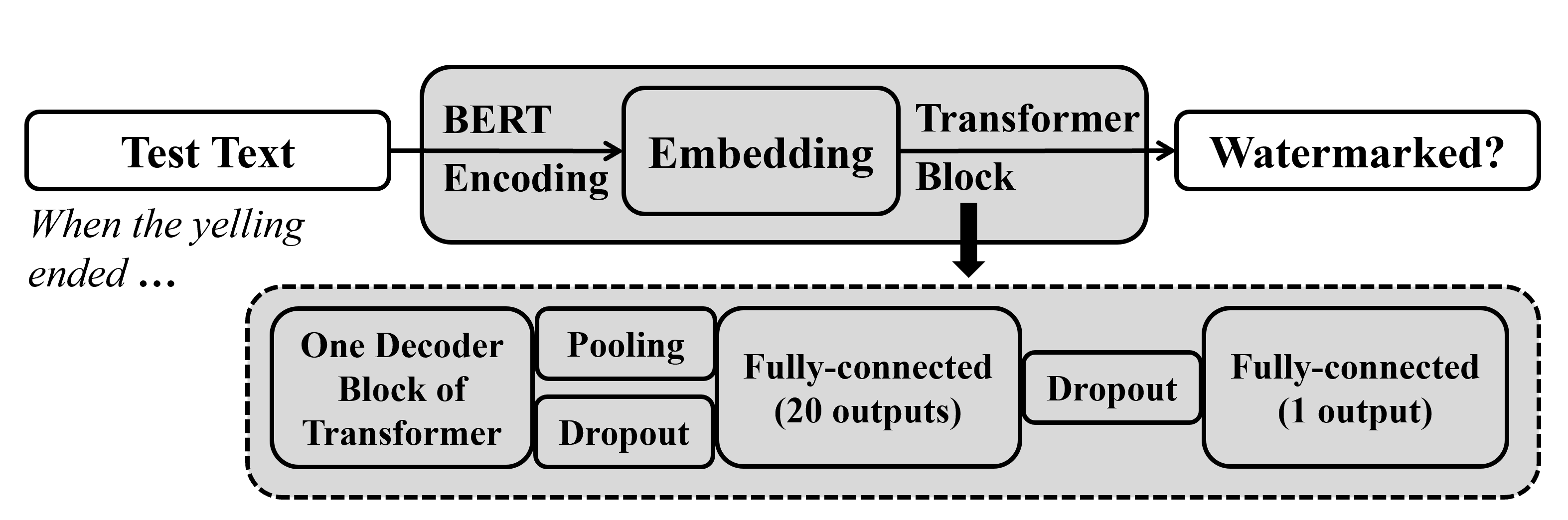}
    \vspace{-0.6em}
    \caption{Watermark detection details.}
    \vspace{-0.6em}
    \label{fig:detection}
\end{figure*}

The watermark detector operates as a binary classifier categorizing inputs into "watermarked" and "unmarked" classes, leveraging network architectures inherent in transformers~\cite{10.5555/3295222.3295349}. We have used the Bidirectional Encoder Representations from Transformers (BERT) pre-trained model which is capable of capturing the contextual meaning of words in a sentence. Unlike traditional methods that treat each word as independent, BERT considers the entire context of the sentence, including the relationships between words. Hence, it will possess the capability to recognize sentence modifications and distinguish between marked and unmarked sentences. Furthermore, BERT serves as a powerful feature extractor, automatically extracting high-dimensional representations of text at various levels of granularity. It's scalability and generalization capabilities enable it to handle diverse datasets and adapt to different domains and languages with minimal additional training. The architecture of this classifier is delineated in Figure~\ref{fig:detection}.

The watermark detection classifier endeavors to minimize the ensuing binary cross-entropy loss:

\begin{equation} \label{eq:loss}
\begin{aligned}
    \mathcal{L} = y_{i} \cdot \log(p(y_{i})) + (1 - y_{i}) \cdot \log(1-p(y_{i})), 
\end{aligned}
\end{equation}

\noindent where $y_{i}$ denotes the label, and $p(y_{i})$ represents the predicted probability. The parameters of the BERT encoder are initially frozen, allowing the loss to converge with the transformer block being trainable. Upon convergence of the loss with a frozen BERT, the parameters of BERT are unfrozen, the learning rate of Adam is attenuated, and training is recommenced until loss convergence is reattained. This iterative training paradigm can precipitate a notable enhancement in prediction performance training solely the transformer block. The outcomes of the training regimen are elaborated in section~\ref{subsec:detectionEval}. This architecture, post convergence, embodies the watermark detector. Given that the detector necessitates no access to the original data for prediction execution, the methodology is characterized as blind.

\subsection{Watermark Detection for Multiple Sentences}
\label{subsec:ProposedSentenceCollection}

A prominent application scenario for the proposed watermarking technique is its deployment on a collection of sentences. Consequently, the classification outcome is contingent on the majority classification rendered for each individual sentence. Employing the binomial distribution, it can be demonstrated that the likelihood of accurately classifying a sentence collection converges to near perfection as the volume of sentences in the collection escalates, provided the probability of accurately classifying a single sentence is reasonably high ($>85\%$). Notwithstanding, a superior probability of correct classification for a single sentence implies a reduced sentence count is requisite to attain near-perfect accuracy. Algorithm~\ref{algo:detection_algo} comprehensively outlines the entire working procedure.

\begin{algorithm}
\caption{Watermark Detection}
\label{algo:detection_algo}
\begin{algorithmic}[1]
\Function{WatermarkDetection}{test\_text}
    \State embeddings $\gets$ \Call{BERT\_Encode}{test\_text}
    \State decoder\_output $\gets$ \Call{TransformerDecoderBlock}{embeddings}
    \State pooled\_output $\gets$ \Call{Pooling}{decoder\_output}
    \State dropout\_output $\gets$ \Call{Dropout}{pooled\_output}
    \State fc\_20\_output $\gets$ \Call{FullyConnected}{dropout\_output, 20}
    \State dropout\_fc $\gets$ \Call{Dropout}{fc\_20\_output}
    \State watermark\_score $\gets$ \Call{FullyConnected}{dropout\_fc, 1}
    \State \Return watermark\_score
\EndFunction
\end{algorithmic}
\end{algorithm}

The proof underpinning this claim is articulated as follows:
Presume the probability of accurately classifying a sentence as watermarked or not is denoted by $p$, and remains consistent across all sentences. 
In a scenario where at least half of the sentences in a text comprising $n$ sentences are accurately classified, the entire text is deemed correctly classified. 
It can be substantiated that the probability of accurately classifying exactly $x$ sentences can be encapsulated by the binomial probability, denoted as $P(x)$. Hence, the probability $P(x > \lceil n/2 \rceil)$ can be formulated as the summation in Equation~\eqref{eq:binomial}:

\vspace{-1.0em}
\begin{equation} \label{eq:binomial}
\begin{aligned}
P(x > \lceil n/2 \rceil) = \sum_{i=\lceil n/2 \rceil}^{n} {n\choose i} \times p^{i} \times (1 - p)^{n - i}.
\end{aligned}
\end{equation}

\section{Experiments}
\label{experiments}

This section illustrates the effectiveness of \mn by analyzing its properties in regard of text watermarking. Dataset preparation is explained in section~\ref{subsec:datasetpreparation}.
The reliability of the watermark detection is shown in section~\ref{subsec:detectionEval}. Section~\ref{subsec:a_s} explains the ablation study.
Section~\ref{subsec:TradeoffEval} provides a summary of the imperceptibility, and the imperceptibility and detection accuracy trade-off. 
Comparisons are made between \mn and traditional text watermarking methods. 
Section~\ref{subsec:robustnessEval} provides an analysis of the experiments used to test robustness, which is followed by an evaluation of the empirically observed runtime in section~\ref{subsec:runtimeEval}. 

\subsection{Dataset}
\label{subsec:datasetpreparation}

\textbf{Training Data:} A dataset comprising 34,489 sentences was assembled from the Dolly ChatGPT Dataset~\cite{dolly_chatgpt}. This approach aims to underscore the generalization capability of the proposed \mn. Robust performance across diverse textual genres exemplifies the model's aptitude for generalizing to arbitrary text. Evaluations have been also conducted on texts engendered by expansive language models such as ChatGPT, as depicted in an instance in Figure~\ref{fig:ChatGPTParagraph}. Within the training set, half of the sentences are watermarked employing the methodology delineated in section~\ref{subsec:EmbeddingMethod}, whilst the remainder are retained unaltered. This yields a dataset encompassing nearly 17,000 watermarked samples and approximately 17,000 unmarked samples. The corpus of watermarked and unmarked sentence samples are randomly amalgamated, with 75\% earmarked for training, and the residual 25\% allocated for validation—this composition underpins the training of the detector. To facilitate the assessment of imperceptibility in section~\ref{subsec:TradeoffEval}, a dataset encapsulating all 34,489 sentences as original and watermarked pairs is retained.

\noindent \textbf{Testing Data:} 
We assessed the performance of our model by subjecting it to testing using C4 datasets~\cite{huggingface-datasets} containing multiple sentences. To evaluate its performance, we systematically extracted 100 tokens at a time, aggregating them into a unified dataset featuring numerous sentences. This process yielded a total of 8,800 datasets. Subsequently, we conducted rigorous testing on these datasets, incorporating both single and multiple synonym substitutions to gauge the model's adaptability and effectiveness.

\subsection{Watermark Detection Accuracy}
\label{subsec:detectionEval}

The proposed watermark detection classifier is trained using the dataset discussed in Section~\ref{subsec:ExtractionMethod}.
We train the architecture with the parameters of the pre-trained BERT encoder frozen for 6 epochs, with an Adam learning rate set to 0.0001.
Then, we unfreeze the pre-trained BERT architecture, reduce the learning rate of Adam~\cite{DBLP:journals/corr/KingmaB14} to 0.000001, and train for 50 more epochs. In our training model, 148 million parameters have been used. The result validation accuracy, which represents the sentence-level detection accuracy on the dolly validation dataset, is 86.52\% for single synonyms and 94.87\% for multiple synonyms. And for C4 datasets, its 76.30\% for single synonyms and 95.72\% for multiple synonyms substitution.

Additionally, we conduct this training process on several versions of the dataset, each with an increasing number of sentences.
We observe that as we continually increase the size of the dataset, the validation accuracy improves.
Training with an increasing number of sentences could further improve the sentence-level prediction accuracy.
We find that the current training is balanced on table~\ref{tbl:AccVsNumSentencesSingle}, table~\ref{tbl:AccVsNumSentencesMultiple} and Section~\ref{subsec:TradeoffEval}, as this validation yields near-perfect prediction accuracy with only a small collection of sentences.

As elucidated by the binomial distribution in Section~\ref{subsec:ProposedSentenceCollection}, the probability of accurately classifying a collection of sentences markedly increases with the augmentation of the sentence count in the text, attributable to our sentence-level insertion process.
Assuming the likelihood of accurately classifying a single sentence aligns closely with the validation accuracy computed during training, and that this likelihood remains consistent across all sentences, we can forecast the probability of accurately classifying a collection of sentences utilizing the summation outlined in Eq.~\eqref{eq:binomial}. 
Under this assumption, the probability of correct prediction corresponding to varying sentence counts is tabulated in Table~\ref{tbl:AccVsNumSentencesSingle} and Table~\ref{tbl:AccVsNumSentencesMultiple}. Table~\ref{tbl:AccVsNumSentencesSingle} and Table~\ref{tbl:AccVsNumSentencesMultiple} underscore the reliability of the method, highlighting an increased likelihood of accurate detection as the number of sentences rises. This trend is observed for both single and multiple synonyms substitution, encompassing both dolly and C4 datasets.

\vspace{-0.3cm}
\begin{table} [H]

\caption{Sentence Count on Detection Accuracy (\%) (Single Synonym)}
\vspace{-0.5em}
\centering
\begin{tabular}{ccc}
\hline
Num Sentences & Dolly & C4 \\
& (\%) & (\%) \\
\hline
1 & 86.52 & 76.30\\
5 & 98.02 & 90.97\\
10 & 99.92 & 98.49\\
20 & 99.99 & 99.75\\
30 & 100.00 & 99.96\\
50 & 100.00 & 99.99\\
60 & 100.00 & 100.00\\
\hline
\end{tabular}
\label{tbl:AccVsNumSentencesSingle}
\end{table}

\vspace{-0.3cm}
\begin{table} [H]

\caption{Sentence Count on Detection Accuracy(\%) (Multiple Synonyms)}
\vspace{-0.5em}
\centering
\begin{tabular}{ccc}
\hline
Num Sentences & Dolly & C4 \\
& (\%) & (\%) \\
\hline
1 & 94.87 & 95.72\\
5 & 99.88 & 99.92\\
10 & 100.00 & 100.00\\
\hline
\end{tabular}
\label{tbl:AccVsNumSentencesMultiple}
\end{table}



\subsection{Ablation Study}
\label{subsec:a_s}

To evaluate the effectiveness of our proposed method, we conducted an ablation study by systematically removing components from our model and observing their impact on performance. Specifically, we conducted four experiments denoted as A, B, C, and D, each representing a variant of our model with varying degrees of complexity. Experiment D, which incorporates all proposed enhancements, achieved the highest accuracy among the tested configurations. This result suggests that the additional components in experiment D contribute positively to the overall performance of the model. Furthermore, by comparing the accuracy of experiment D with those of experiments A, B, and C, as shown in Table~\ref{tbl:ablation}, we can pinpoint the specific contributions of each component to the model's effectiveness. Our findings underscore the importance of the incorporated enhancements and highlight the significance of their inclusion in our proposed approach.

\vspace{-0.3cm}
\begin{table} [H]

\caption{Ablation Study}
\vspace{-0.2em}
\centering
\begin{tabular}{|c|c|c|c|c|}
\hline
Experiment & A & B & C & D \\ \hline
Single Synonyms & \checkmark & \checkmark & & \\ 
Multiple Synonyms & & & \checkmark & \checkmark \\  
Handling Singular/Plural Number & & \checkmark & & \checkmark \\ 
Handling Parts of Speech & & \checkmark & & \checkmark \\ 
Detection Accuracy (\%) & 81.38 & 86.52 & 92.74 & 94.87 \\ 
\hline
\end{tabular}
\label{tbl:ablation}
\end{table}

\subsection{Imperceptibility of Watermark Insertion}
\label{subsec:TradeoffEval}

A sentence bearing an imperceptible watermark should maintain grammatical correctness and retain the same meaning as the original sentence. 
Thus, the imperceptibility of text watermarking should be gauged by sentence meaning similarity. 
The Universal Sentence Encoder~\cite{cer2018universal} encapsulates the semantic meaning of sentences into an embedding vector, enabling the measurement of sentence meaning similarity through the computation of cosine similarity between two sentence embeddings. 
Hence, we propose to quantify the imperceptibility of text watermarking using the Sentence Meaning Similarity ($\text{SMS}$):

\begin{equation}\label{eq:mSMS}
    \text{SMS} = S(\text{encode}(o), \text{encode}(m)), 
\end{equation}

\noindent where $o$ denotes the original text, $m$ denotes the watermarked text, $\text{encode}(\cdot)$ represents a neural network that computes a semantic embedding (e.g., the Universal Sentence Encoder), and $S$ is a function that computes the similarity between the vectors (cosine similarity is utilized in this paper).
Computing the mean $\text{SMS}$ ($m\text{SMS}$) over a dataset provides an average measure of text watermark imperceptibility.
We have performed our experiment for the test dataset discussed in Section~\ref{subsec:ExtractionMethod} and we are able to achieve 0.9765 $m\text{SMS}$ for single synonyms and 0.9892 $m{SMS}$ for multiple synonyms while the traditional method provides 0.9794 $m{SMS}$.
The high $m\text{SMS}$ value exemplifies the imperceptible watermarking of texts by \mn.
An illustration of watermarking a text produced by ChatGPT is presented in Figure~\ref{fig:ChatGPTParagraph}, with additional examples of original and watermarked paragraphs available in the supplementary documents.

\begin{figure}[th!]
    \centering
    \vspace{-1.0em}
    \includegraphics[width=0.55\linewidth]{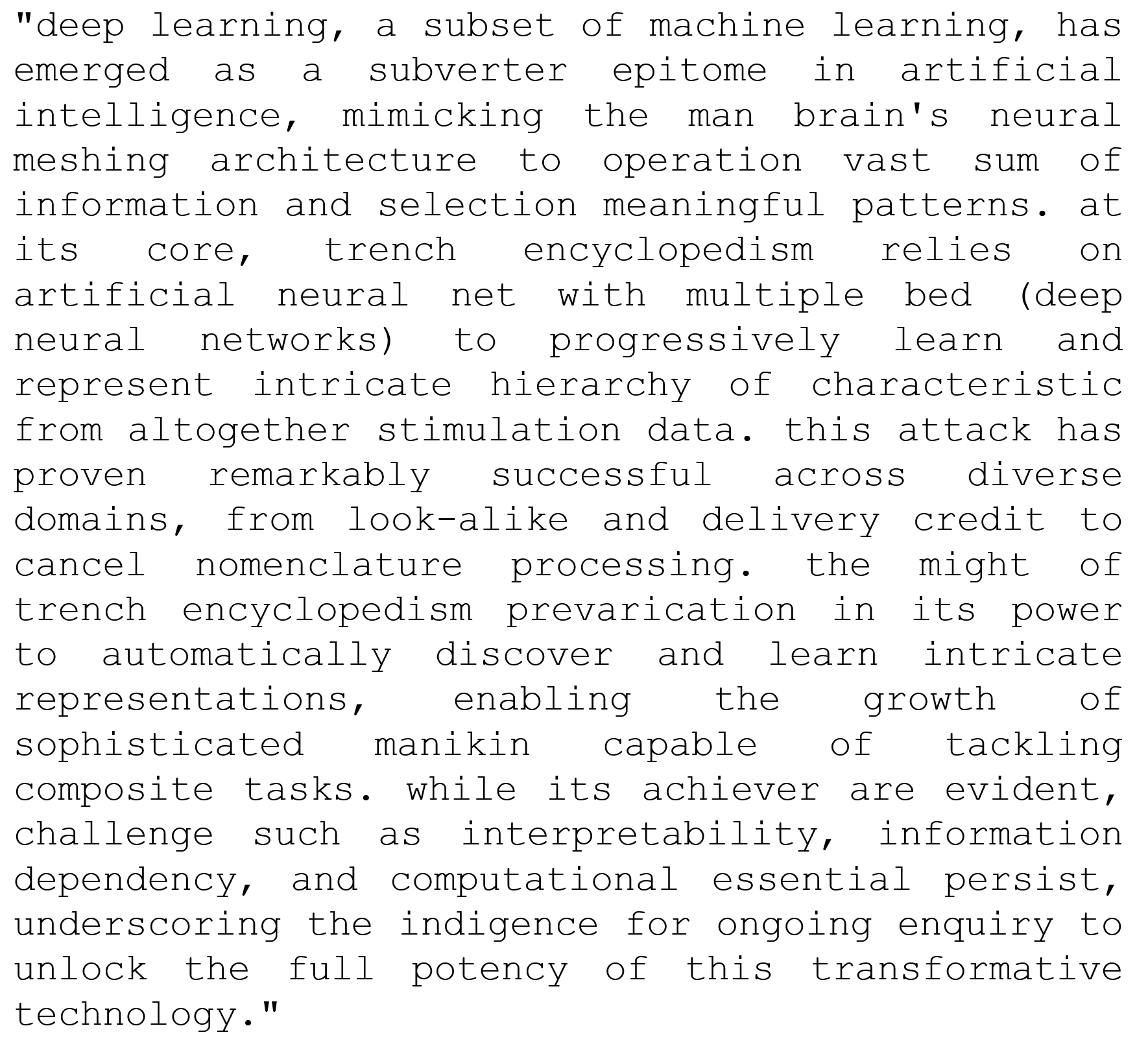}
    \vspace{-1.0em}
    \caption{ A watermarked example from ChatGPT with prompt "Give me a short essay about deep learning". 
    }
    \vspace{-1.5em}
    \label{fig:ChatGPTParagraph}
\end{figure}

For analytical purposes, we implement traditional text watermarking and test the proposed watermark detection network at a single-sentence level.
Specifically, we create an implementation of semantic watermarking using WordNet~\cite{10.1145/219717.219748} to select synonyms.
Although this traditional method achieved some success, the replacement word occasionally rendered the sentence nonsensical, as this method did not account for sentence structure.
Some sentence examples are illustrated in Table~\ref{tbl:exampleSentences} (additional sentence examples can be found in the supplementary documents).
While the traditional method can be effectively detected by our detection network for a single sentence, as depicted in Table~\ref{tbl:QualityVsDetectionTradeoff}, the $mSMS$ on the collected dataset significantly improves with \mn multiple synonyms and not that much far from single synonyms as well.

\begin{table}[H]
\caption{A few example sentences: 1. the original text; 2. watermarked text by the traditional method; 3. watermarked text by the \mn. with single synonym substitution; and 4. watermarked text by the \mn with multiple synonyms substitution}
\vspace{-0.5em}
\centering
\begin{tabular}{|p{7.5cm}|} 
\hline
1. Which episodes of season four of game of thrones did michelle maclaren direct. \\
\hline
2. Which \textbf{installment} of season four of game of thrones did michelle maclaren direct. \\
\hline
3. Which \textbf{sequence} of season four of game of thrones did michelle maclaren direct. \\
\hline
4. Which \textbf{sequence} of season \textbf{quaternity} of game of thrones did michelle maclaren direct. \\ \hline \\ \hline
1. Who saved andromeda from the sea monster? \\
\hline
2. Who saved andromeda from the \textbf{ocean} monster? \\
\hline
3. Who saved andromeda from the \textbf{ocean} monster? \\
\hline
4. Who saved andromeda from the \textbf{ocean monstrosity}? \\ 
\hline
\end{tabular}
\label{tbl:exampleSentences}
\end{table}

\subsection{Robustness}
\label{subsec:robustnessEval}
Robustness in the domain of image watermarking implies that the watermark must remain invariant to malicious attacks or unintentional modifications~\cite{WAN2022226}.
Translating this notion of robustness to text watermarking is fairly straightforward.
A robust text watermarking method should ensure that removing the watermark is challenging, whether the removal attempts are unintentional, arising from normal processing, or intentional attacks targeting the watermark.
For watermark detection to fail, the watermarked text should need to be altered beyond recognition.

Given that this is an emerging area, no standard method exists to measure robustness for text watermarking~\cite{TextWatermarkingReview}.
Therefore, we propose a metric named Mean Impact of Attack ($mIOA$) to measure robustness. The $IOA$ is defined as follows:

\vspace{-1.0em}
\begin{equation}\label{eq:mIOA}
    IOA(x,y) = (1 - |detect(x) - y|) - (1 - |detect(x_{a}) - y|),
\end{equation}
\vspace{-1.0em}

\noindent where $x$ represents the target data (text of one or more sentences in this paper), $x_{a}$ denotes the attacked data obtained by arbitrarily attacking $x$, $y$ signifies the label for $x$ (watermarked or unmarked), and $detect(\cdot)$ denotes the utilization of the detection network to output the predicted label of the input.
$IOA$ gauges the change in accuracy following an attack on the data.
A positive $IOA$ indicates a detrimental effect on prediction performance due to the attack, while a negative $IOA$ indicates improved prediction performance post-attack (which should be rare).
An $IOA$ further from 0 (either less than or greater than) signifies a higher impact from the attack.
An $IOA$ of 0 indicates the attack did not affect the prediction accuracy.
Calculating the mean $IOA$ over a dataset yields the $mIOA$.

\noindent \textbf{Data for Robustness Test.}
We have prepared two sets of data: one with watermarked text and one with unmarked text. 
Each set contains 1000 collections, with each collection comprising 20 sentences. 
These sentences are randomly selected from the testing set described in Section~\ref{subsec:ExtractionMethod}. 

We then define several attacks and compute the $mIOA$ for each attack to gauge the robustness of our watermarking technique.
These attacks are designed to progressively modify the text, with the severity of each attack increasing the dissimilarity between the modified and original texts.
Each attack also represents a common interaction with the text.
By attacking both watermarked and unmarked data, we aim to evaluate the detection accuracy for both types of data, which helps ensure that our system is equally effective at detecting watermarks and identifying unmarked data.

\noindent \textbf{Remove Sentences Attack.} 
We remove a selected number of $n$ sentences from the text. This action reduces the watermark presence in the text, thus challenging the robustness of the detection.
Table~\ref{tbl:RemovedSentAttack} presents the $mIOA$ on both watermarked and unmarked datasets for several values of $n$. 
In all cases, the total number of sentences is 20.

\vspace{-0.75em}
\begin{table} [H]
\caption{Remove Sentences Attack}
\centering
\begin{tabular}{ccc}
\hline
Removed & Watermarked & Unmarked \\
Sentences & $mIOA$ & $mIOA$ \\
\hline
1 & 0.000 & 0.022 \\
3 & 0.000 & 0.033 \\
5 & 0.001 & 0.028 \\
10 & 0.029 & 0.030 \\
15 & 0.058 & 0.169 \\
17 & 0.110 & 0.191 \\
19 & 0.206 & 0.280 \\
\hline
\end{tabular}
\label{tbl:RemovedSentAttack}
\end{table}

The results show that the $mIOA$ increases as the severity of the attack intensifies (i.e., more sentences are removed), yet the performance remains commendable as the $mIOA$ stays close to 0. 
Interestingly, the $mIOA$ is consistently higher on the watermarked data.

\noindent \textbf{Add Sentences Attack.} 
In this attack, a specified number of sentences (represented by $n$) with the opposite label are randomly added to the text. For instance, watermarked sentences are added to unmarked text. Increasing the value of $n$ challenges the robustness of the detection, as it dilutes the percentage of text that corresponds to the expected label. 
Table~\ref{tbl:AddedSentAttack} illustrates the $mIOA$ on the watermarked and unmarked datasets for several values of $n$.

\vspace{-0.75em}
\begin{table} [H]
\caption{Add Sentences Attack}
\vspace{-0.5em}
\centering
\begin{tabular}{ccc}
\hline
Added & Watermarked & Unmarked \\
Sentences & $mIOA$ & $mIOA$ \\
\hline
1 & 0.000 & 0.026 \\
3 & 0.001 & 0.061 \\
5 & 0.003 & 0.115 \\
10 & 0.069 & 0.269 \\
\hline
\end{tabular}
\label{tbl:AddedSentAttack}
\end{table}

The data shows that the $mIOA$ increases as $n$ increases. 
\mn maintains a high performance, as the $mIOA$ remains close to 0 for a reasonable $n$. 

\noindent \textbf{Replace Sentences Attack.}
This attack adopts a similar data dilution approach as the add sentences attack. 
It distorts the text data by replacing $n$ existing sentences in the text with randomly selected sentences of the opposite type, where $n$ is a specified integer. Table~\ref{tbl:ReplacedSentAttack} presents the watermarked and unmarked $mIOA$ for several values of $n$. 

\vspace{-0.1cm}
\begin{table} [H]
\caption{Replace Sentences Attack}
\centering
\begin{tabular}{ccc}
\hline
Replaced & Watermarked & Unmarked \\
Sentences & $mIOA$ & $mIOA$ \\
\hline
1 & 0.001 & 0.013 \\
3 & 0.008 & 0.058 \\
5 & 0.040 & 0.126 \\
7 & 0.103 & 0.210 \\
\hline
\end{tabular}
\label{tbl:ReplacedSentAttack}
\end{table}

The $mIOA$ increases as $n$ increases, and \mn remains close to 0, indicating a minimal impact on detection performance.
Since the modified text becomes increasingly dissimilar to the original text post-attack, an escalating performance impact is expected and acceptable as the severity of each attack intensifies. 
These experiments affirm that \mn is robust to text modifications stemming from common text interactions.

\subsection{Comparative Analysis}
\label{subsec:ComparisonEval}
For comparative analysis, we have used individual three methods and their combinations. Thus we have conducted comprehensive experiments on three combinations considering each of these methods as indicated in table~\ref{tbl:c_analysis}. We perform the comparison in terms of imperceptibility and detection accuracy which has been shown in table~\ref{tbl:QualityVsDetectionTradeoff}

\vspace{-0.3cm}
\begin{table} [H]

\caption{Summary of the experiments conducted.}
\vspace{-0.5em}
\centering
\begin{tabular}{|p{1.5cm}|p{1cm}|p{1cm}|p{1.5cm}|p{1.5cm}|}
\hline
     Experiment&Single Word Change&Multiple Word Change &Handling Singular/Plural Number 
     &Handling Parts of Speech \\
     \hline
     Existing&\checkmark&&&\\
     \hline
     Single Synonyms&\checkmark&&\checkmark&\checkmark\\
     \hline
     Multiple Synonyms&&\checkmark&\checkmark&\checkmark\\
     \hline
\end{tabular}
\label{tbl:c_analysis}
\end{table}

\vspace{-0.5em}
\begin{table} [H]
\caption{Comparative analysis in terms of $mSMS$ and Detection Accuracy}
\vspace{-0.5em}
\centering
\begin{tabular}{ccc}
\hline
Method & $mSMS$ & Detection Accuracy \\
\hline
\mn (Single Synonyms) & 0.9765 & 0.8652 \\
\mn (Multiple Synonyms) & 0.9892 & 0.9487\\
Traditional & 0.9794 & 0.8836 \\
\hline
\end{tabular}
\label{tbl:QualityVsDetectionTradeoff}
\end{table}
\vspace{-1.0em}

From table~\ref{tbl:QualityVsDetectionTradeoff} we can see that our DeepTextMark with multiple synonyms performed very well in terms of both imperceptibility and detection accuracy.

While the domain of content watermarking in LLMs outputs is burgeoning, the variety of methodologies developed thus far remains limited. We have performed a comparative analysis between our approach and the method proposed by John \textit{et al}.~\cite{kirchenbauer2023watermark}. For clarity within the context of this paper, we will refer to their method as the Watermark Logit Processor (WLP) method, to prevent any naming confusion. It's important to highlight that the WLP method necessitates access to LLMs, specifically utilizing them as a logit processor to favor the selection of "green" tokens during text generation. On the other hand, our proposed method operates independently and does not require access to LLMs.

\begin{figure}[ht!]
    \centering
    \vspace{-0.6em}
    \includegraphics[width=0.50\linewidth]{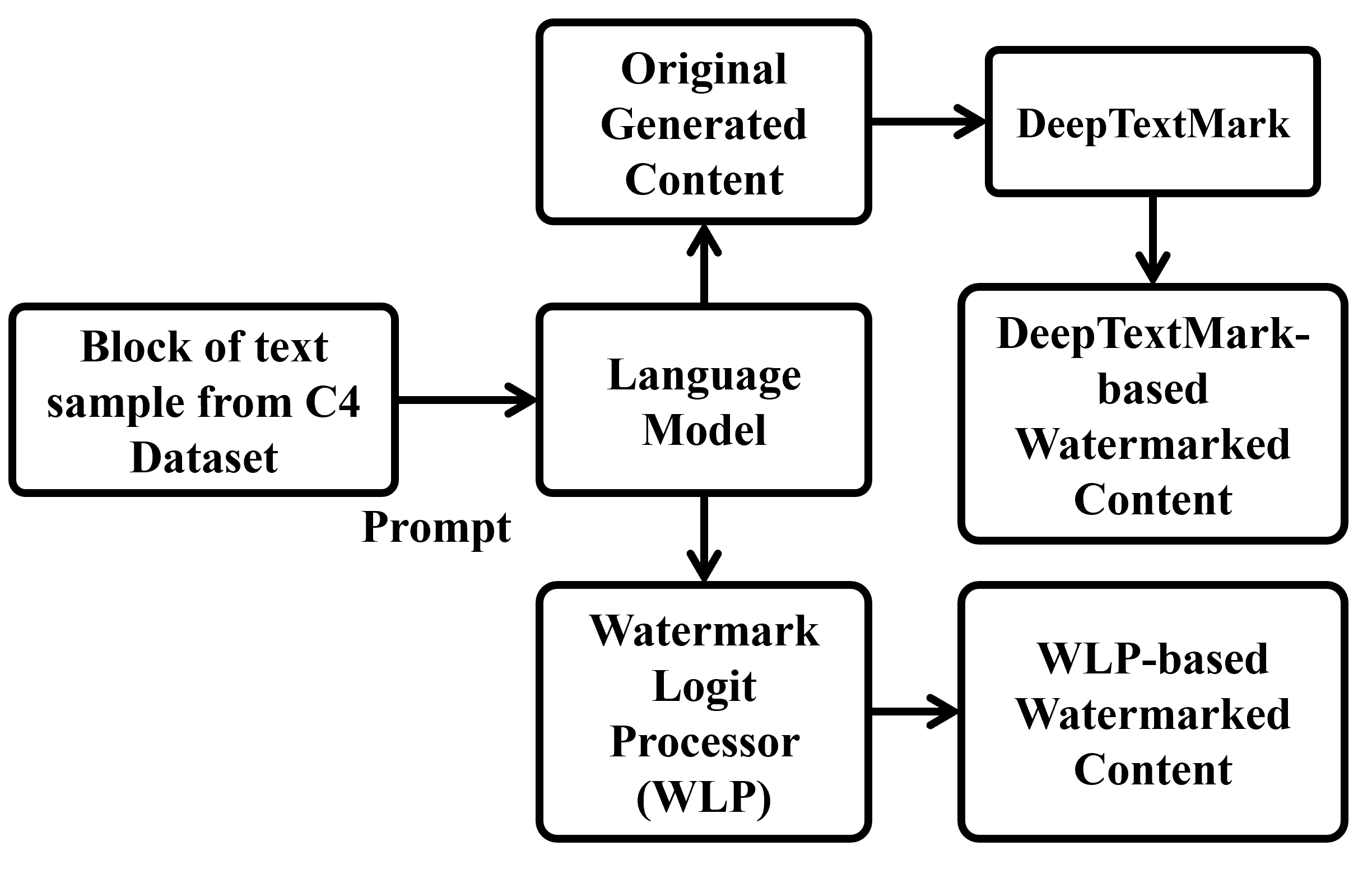}
    \vspace{-0.6em}
    \caption{Sample generation process for testing and comparing watermark detection accuracy of \mn and WLP.} 
    \vspace{-0.6em}
    \label{fig:sample_generation}
\end{figure}

To ensure a fair comparison, it is imperative that both methods are evaluated using the same source of text, specifically an LLM. Consequently, for text generation, we have employed the Open Pre-trained Transformer (OPT-2.7B). The primary objective of this experiment is to apply our method, alongside the WLP method, to watermark the content generated by OPT-2.7B and subsequently evaluate the detection accuracy for comparison purposes. To generate a substantial amount of text content, we utilized a subset of the C4 dataset, comprising 22k text samples, as the source of prompts for the LLM in a seeded environment to yield deterministic outputs with a set of 500 sequences of length T = 200 token sequences which is similar to the WLP paper. The authors of WLP papers proposed two different methods which we denote \textbf{WLP-multinomial sampling} and \textbf{WLP-beam search} to avoid confusion. 
With this setup, upon inputting text (prompt) samples from the C4 dataset into the base LLM, we obtain blocks of text, which we term the "Original Generated Content." Subsequently, we apply our proposed method to the "Original Generated Content" to produce the DeepTextMark-based watermarked content. Conversely, when we incorporate the WLP logit processor with the base LLM, the identical input text samples yield the WLP Watermarked Content. Figure~\ref{fig:sample_generation} illustrates the text generation methodology employed for the comparative evaluation between DeepTextMark and WLP. 
In this configuration, our model demonstrates a notable detection rate of 90.66\%. This outcome, achieved despite training on the distinct Dolly dataset, underlines the robust generalization capability of our approach, affirming its effectiveness across diverse datasets.

In our experimental evaluation, we utilized a subset of 500 data points to assess the watermark detection performance of both models. Despite the distinct datasets employed in training our model \mn, it demonstrates a commendable detection rate, only marginally lower than that reported in the WLP paper. Specifically, DeepTextMark achieved an accuracy of 90.74\%, closely approaching the 92.42\% accuracy of the WLP model. This proximity in performance is noteworthy, considering the differences in training datasets. Table~\ref{tab:detection_rate_comparison} presents a detailed comparative analysis of the watermark detection accuracies between DeepTextMark and WLP.


\begin{table}[ht!]
    \centering
    \caption{Detection accuracy of \mn and WLP with smaller datasets}
    \begin{tabular}{cc}
    \hline
        Model & Detection Accuracy (\%) \\
        \hline
        WLP & 92.43 \\
        \mn & 90.74 \\
    \hline
    \end{tabular}
    \label{tab:detection_rate_comparison}
\end{table}

We conducted a robustness comparison between the two models, considering three attack types: text insertion, deletion, and substitution. Text insertion attacks add extra tokens post-generation, while text deletion removes tokens from the generated output, potentially diminishing text quality by reducing the effective language model (LM) context width. Text substitution attacks involve replacing one token with another, which can be automated through dictionary or LM techniques but may degrade text quality.

Our comparative analysis, summarized in Table~\ref{tab:robustness_comparison}, reveals the robustness of \mn and WLP. The WLP study involved meticulous parameter adjustments to optimize their model's performance. Despite being trained on the Dolly Dataset, our model exhibited superior performance when tested on the C4 dataset produced by the LLM, outperforming in most scenarios for watermark detection accuracy. For robustness evaluation, we introduced new metrics while also using the True Positive Rate (TPR) and False Negative Rate (FNR) metrics from the WLP paper to ensure a fair assessment.

The Area Under the Receiver Operating Characteristic (AUC) curve and True Positive Rate (TPR) are key metrics in binary classification. AUC illustrates the trade-off between sensitivity (TPR) and 1 - specificity (False Positive Rate) across different thresholds, ranging from 0 to 1. A value of 0.5 implies no discriminative ability, whereas 1 indicates perfect classification. Higher AUC values denote superior model performance.
TPR, or sensitivity/recall, is the ratio of correctly identified positive instances to all actual positives, defined as:
\( \text{TPR} = True Positives / (True Positives + False Negatives) \). 
Conversely, the False Negative Rate (FNR) quantifies the proportion of positives incorrectly classified as negatives:
\( \text{FNR} = False Negatives / (Positives + False Negatives) \)

A superior TPR, signifying \mn's proficiency in correctly identifying positive instances while minimizing false negatives underscores its efficacy in capturing the majority of actual positive cases. Concurrently, the smaller FNR suggests a reduced probability of overlooking positive instances, highlighting \mn's competence in averting false negatives and precisely identifying positive cases. In light of our model's outperformance compared to WLP, it can be inferred that \mn demonstrates a heightened capability in detecting watermarked sentences, surpassing the performance of WLP in this regard. This substantiates the conclusion that our model excels in discerning watermarked content more effectively.

\begin{table}[ht!]
    \centering
    \caption{Robustness comparison of \mn and WLP}
    \begin{tabular}{cccc}
    \hline
        Model & $\epsilon$ & TPR & FNR\\
        \hline
        multinomial sampling & 0.1 & 0.819 & 0.181 \\
        multinomial sampling & 0.3 & 0.353 & 0.647 \\
        multinomial sampling & 0.5 & 0.094 & 0.906 \\
        multinomial sampling & 0.7 & 0.039 & 0.961 \\
        beam search & 0.1 & 0.834 & 0.166 \\
        beam search & 0.3 & 0.652 & 0.348 \\
        beam search & 0.5 & 0.464 & 0.536 \\
        beam search & 0.7 & 0.299 & 0.701 \\
        \mn & - & 0.830 & 0.170 \\
    \hline
    \end{tabular}
    \label{tab:robustness_comparison}
\end{table}

\begin{figure}[ht!]
    \centering
    \vspace{-0.6em}
    \includegraphics[width=0.5\linewidth]{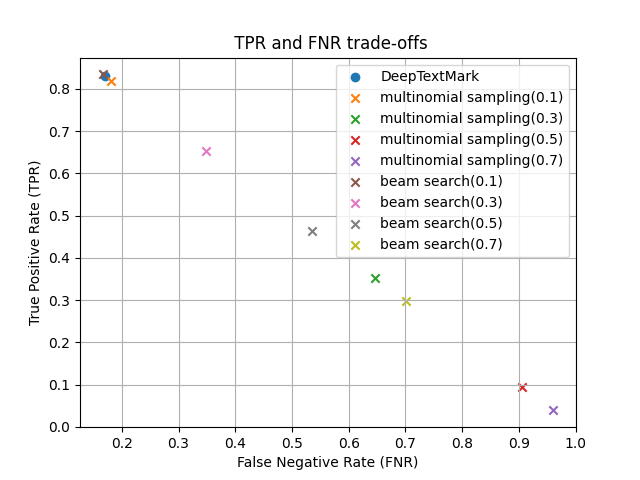}
    \vspace{-0.6em}
    \caption{TPR and FNR Trade-offs} 
    \vspace{-0.6em}
    \label{fig:trade_off}
\end{figure}

Figure~\ref{fig:trade_off} delineates the interplay between TPR and FNR for our proposed method about the established method, WLP. Each data point on the plot encapsulates the performance of a method at distinct decision thresholds. The visual examination of the scatter plot underscores that our method does not lag behind WLP in terms of TPR and FNR characteristics across various operational points. This observation is crucial in establishing the efficacy of our method, aligning it favorably with the performance benchmarks set by WLP.

\subsection{Emperical Running Speed}
\label{subsec:runtimeEval}
This section evaluates the running speed of \mn.
The experiments concerning running speed are conducted on an Intel i9-13900k CPU.
We measure the time taken for watermark insertion across 1000 unmarked sentences and compute the sentence-level average watermark insertion time.
Similarly, we time the watermark detection process on 1000 watermarked sentences and compute the average detection time.
The average times for watermark insertion and detection, in seconds, are provided in Table~\ref{tbl:DeepTextMarkRuntime}.

\vspace{-0.1cm}
\begin{table} [H]
\caption{\mn Runtime on a single CPU core}
\centering
\begin{tabular}{ccc}
\hline
Component & Time per Sentence (seconds) \\
\hline
Insertion & 0.27931 \\
Detection & 0.00188 \\
\hline
\end{tabular}
\label{tbl:DeepTextMarkRuntime}
\end{table}

As demonstrated, both the insertion and detection processes run quickly, serving as efficient "add-on" components for text source detection.
The insertion process incurs a higher overhead compared to the detection process.
It is important to note that these experiments were conducted using only a single core of the CPU.
By parallelizing the implementation, the overhead from the insertion process could be significantly reduced, especially on server-level machines, which are typically employed to implement LLMs in our target application scenario.

\section{Conclusion}
\label{conclusion}
Recently, the use of LLMs has surged significantly in both industry and academia, mainly for text generation tasks.
Nevertheless, in certain scenarios, it is crucial to ascertain the source of text—whether it is generated by an LLM or crafted by a human.
Addressing this requirement, we introduce a deep learning-based watermarking technique designed for text source identification, which can seamlessly integrate with existing LLM-driven text generators.
Our proposed method, \mn, stands out due to its blind, robust, reliable, automatic, and imperceptible characteristics.
Unlike common direct classification techniques~\cite{gptzero} for source detection that demand a substantial amount of characters for accurate prediction, our watermarking technique enables both watermark insertion and detection at the sentence level.
Our findings demonstrate that with the insertion of watermarks, the accuracy of our detection classifier can approach near-perfection with merely a small set of sentences.
Given that the watermark is embedded in each sentence individually, the robustness and reliability of the watermark enhance with an increasing number of sentences.
The core advantages of our work include: an "add-on" text watermarking method facilitating the detection of generated text without requiring access to the LLMs' generation phase; an automatic and imperceptible method for watermark insertion; and a robust, high-accuracy, deep learning-based text watermark detection methodology.

While DeepTextMark introduces a significant advancement in text watermarking using deep learning, we recognize a few areas where future enhancements could be beneficial. First, the effectiveness of DeepTextMark is closely tied to the representativeness of the training data. Efforts to diversify this data could further improve its applicability across various text styles and languages. Second, as DeepTextMark functions in a 'plug-in' manner, its utility is contingent on the initial watermarking of the generated text. Without pre-watermarking, detection capabilities are limited, pointing to a dependency that may affect its applicability in certain scenarios. Lastly, while the method currently shows promising results in watermarking texts of standard lengths, we are exploring ways to adapt it more effectively for very short or stylistically diverse texts. These limitations represent opportunities for ongoing research and underscore the potential for continuous improvement in the field of AI-driven text watermarking.

In conclusion, our study has successfully introduced DeepTextMark, a novel deep learning-driven approach for text watermarking, offering a robust solution for distinguishing between human-authored texts and those generated by large language models. As we look toward the future, several promising directions can further enhance and expand the utility of our approach.
We envision enhancing the robustness of DeepTextMark against more advanced text manipulation techniques, especially those using AI-based rewriting tools, to maintain its effectiveness in increasingly sophisticated digital environments. Moreover, exploring scalability to manage larger and more diverse datasets will be crucial in adapting our method for big data applications.
Another significant direction involves extending the compatibility of DeepTextMark with various large language models, broadening its applicability across different AI-generated text scenarios. Developing real-time applications, such as content management system plugins, will also be pivotal in dynamically detecting and managing AI-generated content.
Lastly, we acknowledge the importance of addressing the ethical and legal implications surrounding text watermarking, particularly in terms of privacy and data security in the age of AI. This aspect is critical to ensuring that our methodologies align with societal norms and legal standards.
As we continue to build upon the foundation laid by DeepTextMark, these future endeavors will undoubtedly contribute to the evolving landscape of text watermarking and AI-generated content detection, reinforcing the importance of authenticity and integrity in digital communications.

\bibliographystyle{IEEEtran}
\bibliography{references}

\newpage
\clearpage
\setcounter{page}{1}
{\Large SUPPLEMENTARY DOCUMENTS}

{\Large SENTENCE LEVEL ATTACKS}

In the paper, we presented several attacks on multiple sentences to demonstrate the robustness of sentence collections. Here, we discuss
similar attacks on the tokens of single sentences to demonstrate sentence-level robustness. We prepare another two sets of watermarked and unmarked data respectively. Each set contains 1000 randomly selected sentences from the testing set described in the paper.
\textbf{Remove Tokens Attack:} We remove a chosen number of $n$ tokens randomly from the sentence similar to the remove sentences attack in the paper. Removing tokens distorts the original watermarked text, and has a chance to remove the watermark. Table~\ref{tbl:RemovedTokenAttack} shows the $mIOA$ on the watermarked and unmarked datasets for several values of $n$.

\vspace{-0.75em}
\begin{table} [H]
\caption{Remove Tokens Attack}
\centering
\begin{tabular}{ccc}
\hline
Removed & Watermarked & Unmarked \\
Sentences & $mIOA$ & $mIOA$ \\
\hline
1 & 0.007 & 0.011 \\
2 & 0.033 & 0.055 \\
3 & 0.045 & 0.077 \\
\hline
\end{tabular}
\label{tbl:RemovedTokenAttack}
\end{table}

\textbf{Add Tokens Attack:} We add $n$ random words to a random location in the sentence to perform the add tokens attack, where $n$ is a chosen integer. The source of the random words is the NLTK Words corpus. The add token attack challenges the robustness because the sentence is distorted from its original state by adding noise. Table~\ref{tbl:AddTokenAttack} shows the $mIOA$ on the watermarked and unmarked
datasets for several values of $n$.

\vspace{-0.75em}
\begin{table} [H]
\caption{Remove Tokens Attack}
\centering
\begin{tabular}{ccc}
\hline
Removed & Watermarked & Unmarked \\
Sentences & $mIOA$ & $mIOA$ \\
\hline
1 & -0.005 & 0.050 \\
2 & -0.008 & 0.136 \\
3 & -0.006 & 0.230 \\
\hline
\end{tabular}
\label{tbl:AddTokenAttack}
\end{table}

\textbf{Replace Tokens Attack:} We replace $n$ random words in the sentence with words randomly selected from the NLTK Words corpus, where $n$ is a chosen integer. The replace tokens attack is the
most severe of these sentence-level attacks it is a combination of removing tokens, which may remove the watermark and filling the sentence with noise. Table~\ref{tbl:ReplaceTokenAttack} shows the $mIOA$ on the watermarked
and unmarked datasets for several values of $n$

\vspace{-0.75em}
\begin{table} [H]
\caption{Remove Tokens Attack}
\centering
\begin{tabular}{ccc}
\hline
Removed & Watermarked & Unmarked \\
Sentences & $mIOA$ & $mIOA$ \\
\hline
1 & 0.005 & 0.085 \\
2 & 0.005 & 0.022 \\
3 & -0.005 & 0.331 \\
\hline
\end{tabular}
\label{tbl:ReplaceTokenAttack}
\end{table}

{\Large MORE APPLICATIONS} \\ \\
{\LARGE Pre-detection to Improve Imperceptibility
} \\

During the watermarking process on 1 or more sentences, we can pre-detect the watermark of each sentence to determine whether or not the sentence-level watermark is detectable. If the watermark detection does not succeed, we can instead use the original sentence to improve imperceptibility at the cost of some runtime performance. When watermarking using the pre-detection method, the
watermarker verifies that at least 51\% of the text is watermarkable, otherwise, the given text is determined to not be watermarkable. As the detection is checked ahead of time, reducing the number of watermarked sentences in this manner does not impact downstream detection performance. Additionally, this method could inform users in advance whether or not the given text is watermarkable and the percentage of the text that is detectable.

Consider the following text: \\

\indent Biodiversity, or the variety of life on earth, is essential for the functioning of ecosystems. Ecosystems are complex networks of living organisms and their physical environment, and they rely on biodiversity to maintain their balance and resilience. One of the key benefit of biodiversity is that it supports the provision of ecosystem services such as food, fuel, and water. For example, a diverse array of plants, animals, and microorganisms is necessary for healthy soil, which in turn supports agriculture and the production of food. Biodiversity also plays a crucial role in regulating the earth’s climate. Plants and trees absorb carbon dioxide, a greenhouse gas that contributes to global warming, through the process of photosynthesis. The loss of biodiversity can lead to a reduction in the number of plants and trees, which can then lead to an increase in the concentration of carbon dioxide in the atmosphere. In addition, biodiversity has important cultural and aesthetic values. Many cultures have deep spiritual and social connections to the natural world and its biodiversity and rely on it for their livelihoods and well-being. Biodiversity also provides recreational opportunities such as bird watching, hiking, and fishing, which contribute to physical and mental health. Despite its importance, biodiversity is threatened by a range of human activities such as habitat destruction, overfishing, pollution, and climate change. These threats can have cascading effects on ecosystems and the services they provide. To protect biodiversity, it is necessary to take actions such as protecting and restoring habitats, managing natural resources sustainably, and reducing greenhouse gas emissions. In conclusion, biodiversity is a fundamental component of healthy and functioning ecosystems, providing essential services to humans and the planet. It is our responsibility to protect and conserve biodiversity for present and future generations. \\

When watermarked using DeepTextMark without pre-prediction, the following is the result: \\

\indent Biodiversity, or the variety of life on the planet, is essential for the functioning of ecosystems. Ecosystems are complex networks of living organisms and their physical environment, and they rely on biodiversity to maintain their balance and resilience. One of the key benefits of biodiversity is that it supports the provision of ecosystem services such as food, fuel, and water. For example, a diverse array of plants, animals, and microorganisms is necessary for healthy soil, which in turn supports agriculture and the production of food. Biodiversity also plays a crucial role in regulating the earth’s climate. Plants and trees absorb carbon dioxide, a greenhouse gas that contributes to global warming, through the process of photosynthesis. The loss of biodiversity can lead to a reduction in the number of plants and trees, which can then lead to an increase in the concentration of carbon dioxide in the atmosphere. In addition, biodiversity has important cultural and aesthetic values. Many cultures have deep spiritual and social connections to the natural world and its biodiversity and rely on it for their \textbf{livelihood} and well-being. Biodiversity also provides recreational opportunities such as bird watching, hiking, and fishing, which contribute to physical and mental health. Despite its \textbf{important}, biodiversity is threatened by a range of human activities such as habitat destruction, overfishing, pollution, and climate change. These threats can have cascading effects on ecosystems and the services they provide. To protect biodiversity, it is necessary to take actions such as protecting and restoring habitats, managing natural resources sustainably, and reducing greenhouse gas emissions. In conclusion, biodiversity is a fundamental component of healthy and functioning ecosystems and provides essential services to humans and the planet. It is our responsibility to protect and conserve biodiversity for present and future generations. \\

And, when watermarked using DeepTextMark with pre-prediction, the following is the result: \\

\indent Biodiversity, or the variety of life on the planet, is essential for the functioning of ecosystems. Ecosystems are complex networks of living organisms and their physical environment, and they rely on biodiversity to maintain their balance and resilience. One of the key benefits of biodiversity is that it supports the provision of ecosystem services such as food, fuel, and water. For example, a diverse array of plants, animals, and microorganisms is necessary for healthy soil, which in turn supports agriculture and the production of food. Biodiversity also plays a crucial role in regulating the earth’s climate. Plants and trees absorb carbon dioxide, a greenhouse gas that contributes to global warming, through the process of photosynthesis. The loss of biodiversity can lead to a reduction in the number of plants and trees, which can then lead to an increase in the concentration of carbon dioxide in the atmosphere. In addition, biodiversity has important cultural and aesthetic values. Many cultures have deep spiritual and social connections to the natural world and its biodiversity and rely on it for their \textbf{livelihoods} and well-being. Biodiversity also provides recreational opportunities such as bird watching, hiking, and fishing, which contribute to physical and mental health. Despite its \textbf{importance}, biodiversity is threatened by a range of human activities such as habitat destruction, overfishing, pollution, and climate change. These threats can have cascading effects on ecosystems and the services they provide. To protect biodiversity, it is necessary to take actions such as protecting and restoring habitats, managing natural resources sustainably, and reducing greenhouse gas emissions. In conclusion, biodiversity is a fundamental component of healthy and functioning ecosystems and provides essential services to humans and the planet. It is our responsibility to protect and conserve biodiversity for present and future generations. \\

The two differences in the text are highlighted in bold. Of this watermarked text, 87\% of the sentences are detectable using the
detector. Therefore, the two sentences that are not detectable do not need to be modified. As shown by the differences in the text,
the imperceptibility of the watermark improves when the text is watermarked using the pre-detection method while not impacting detection performance. \\

{\large Detecting Watermarked Percentage} \\

DeepTextMark can be applied to detect the percentage of text that is watermarked. Since the detection occurs at the sentence level,
the detector can produce a count of the sentences that are predicted watermarked. Then, the percentage of watermarked text is computed as the number of watermarked sentences divided by the total number of sentences. For example, consider the following text: \\

Deep learning is a subfield of machine learning that focuses on training artificial neural networks with multiple layers. The term deep refers to the depth of the network, which can range from a few layers to dozens or even thousands of layers. One of the key advantages of deep learning is its ability to automatically discover relevant features and patterns in the data, without the need for manual feature engineering. \\ 

The first sentence is not marked, and the second two sentences are watermarked. This text is then input into the detector to determine the percentage watermarked. The result is 67\%, which is correct.

Applying DeepTextMark in this way can further reduce the impact of attacks discussed in the robustness section of the paper,
as the percentage only considers sentence predictions. Additionally, the percentage detected watermarked could be provided as a
certainty score. \\

{\large ORIGINAL VS TRADITIONAL VS DEEPTEXTMARK} \\

This section provides more comparative examples between the traditional method and DeepTextMark. In the following examples: 1. the original text; 2. watermarked text by the traditional method; 3. watermarked text by the DeepTextMark. with single synonym substitution; and 4. watermarked text by the DeepTextMark with multiple synonyms substitution \\

\begin{enumerate}
\item Virgin australia the trading name of virgin australia airlines pty ltd is an australianbased airline it is the largest airline by fleet size to use the virgin brand it commenced services on 31 august 2000 as virgin blue with two aircraft on a single route it suddenly found itself as a major airline in australia's domestic market after the collapse of ansett australia in september 2001 the airline has since grown to directly serve 32 cities in australia from hubs in brisbane melbourne and sydney
\item Virgo australia the trading name of virgin australia airlines pty ltd is an australianbased airline it is the largest airline by fleet size to use the virgin brand it commenced services on 31 august 2000 as virgin blue with two aircraft on a single route it suddenly found itself as a major airline in australia's domestic market after the collapse of ansett australia in september 2001 the airline has since grown to directly serve 32 cities in australia from hubs in brisbane melbourne and sydney
\item Virgo australia the trading name of virgin australia airlines pty ltd is an australianbased airline it is the largest airline by fleet size to use the virgin brand it commenced services on 31 august 2000 as virgin blue with two aircraft on a single route it suddenly found itself as a major airline in australia's domestic market after the collapse of ansett australia in september 2001 the airline has since grown to directly serve 32 cities in australia from hubs in brisbane melbourne and sydney
\item Virgo australia the trading epithet of virgo australia airlines pty ltd is an australianbased airway it is the largest airway by fleet sizing to utilization the virgo blade it commenced serving on 31 aug 2000 as virgo blueing with 2 aircraft on a one path it suddenly found itself as a major airway in australia's domestic mart after the prostration of ansett australia in sep 2001 the airway has since grown to directly service 32 metropolis in australia from hubs in brisbane melbourne and sydney
\end{enumerate}

\begin{enumerate}
\item Which is a species of fish tope or rope
\item Which is angstrom species of fish tope or rope
\item Which is a mintage of fish tope or rope
\item Which is a mintage of pisces stupa or rophy
\end{enumerate} 

\begin{enumerate}
\item Why can camels survive for long without water
\item Wherefore can camels survive for long without water
\item Why can camels survive for long without urine
\item Why can camels survive for long without urine
\end{enumerate} 

\begin{enumerate}
\item Camels use the fat in their humps to keep them filled with energy and hydration for long periods of time
\item Camel use the fat in their humps to keep them filled with energy and hydration for long periods of time
\item Camels utilization the fat in their humps to keep them filled with energy and hydration for long periods of time
\item Camels utilization the blubber in their gibbousness to living them filled with get-up-and-go and hydration for long menstruation of sentence
\end{enumerate}

\begin{enumerate}
\item Alice's parents have three daughters amy jessy and whats the name of the third daughter
\item Alice's parent have three daughters amy jessy and whats the name of the third daughter
\item Alice's parent have three daughters amy jessy and whats the name of the third daughter
\item Alice's parent have 3 daughter amy jessy and whats the epithet of the tierce girl
\end{enumerate}

\begin{enumerate}
\item The name of the third daughter is alice
\item The epithet of the third daughter is alice
\item The epithet of the third daughter is alice
\item The epithet of the tierce girl is alice
\end{enumerate}

\begin{enumerate}
\item When was tomoaki komorida born
\item When washington tomoaki komorida born
\item When was tomoaki komorida have
\item When was tomoaki komorida born
\end{enumerate}

\begin{enumerate}
\item Komorida was born in kumamoto prefecture on july 10 1981 after graduating from high school he joined the j1 league club avispa fukuoka in 2000 although he debuted as a midfielder in 2001 he did not play much and the club was relegated to the j2 league at the end of the 2001 season in 2002 he moved to the j2 club oita trinita he became a regular player as a defensive midfielder and the club won the championship in 2002 and was promoted in 2003 he played many matches until 2005 in september 2005 he moved to the j2 club montedio yamagata in 2006 he moved to the j2 club vissel kobe although he became a regular player as a defensive midfielder his gradually was played less during the summer in 2007 he moved to the japan football league club rosso kumamoto later roasso kumamoto based in his local region he played as a regular player and the club was promoted to j2 in 2008 although he did not play as much he still played in many matches in 2010 he moved to indonesia and joined persela lamongan in july 2010 he returned to japan and joined the j2 club giravanz kitakyushu he played often as a defensive midfielder and center back until 2012 when he retired
\item Komorida washington born in kumamoto prefecture on july 10 1981 after graduating from high school he joined the j1 league club avispa fukuoka in 2000 although he debuted as a midfielder in 2001 he did not play much and the club was relegated to the j2 league at the end of the 2001 season in 2002 he moved to the j2 club oita trinita he became a regular player as a defensive midfielder and the club won the championship in 2002 and was promoted in 2003 he played many matches until 2005 in september 2005 he moved to the j2 club montedio yamagata in 2006 he moved to the j2 club vissel kobe although he became a regular player as a defensive midfielder his gradually was played less during the summer in 2007 he moved to the japan football league club rosso kumamoto later roasso kumamoto based in his local region he played as a regular player and the club was promoted to j2 in 2008 although he did not play as much he still played in many matches in 2010 he moved to indonesia and joined persela lamongan in july 2010 he returned to japan and joined the j2 club giravanz kitakyushu he played often as a defensive midfielder and center back until 2012 when he retired
\item Komorida was born in kumamoto prefecture on july x 1981 after graduating from high school he joined the j1 league club avispa fukuoka in 2000 although he debuted as a midfielder in 2001 he did not play much and the club was relegated to the j2 league at the end of the 2001 season in 2002 he moved to the j2 club oita trinita he became a regular player as a defensive midfielder and the club won the championship in 2002 and was promoted in 2003 he played many matches until 2005 in september 2005 he moved to the j2 club montedio yamagata in 2006 he moved to the j2 club vissel kobe although he became a regular player as a defensive midfielder his gradually was played less during the summer in 2007 he moved to the japan football league club rosso kumamoto later roasso kumamoto based in his local region he played as a regular player and the club was promoted to j2 in 2008 although he did not play as much he still played in many matches in 2010 he moved to indonesia and joined persela lamongan in july 2010 he returned to japan and joined the j2 club giravanz kitakyushu he played often as a defensive midfielder and center back until 2012 when he retired
\item Komorida was born in kumamoto prefecture on july x 1981 after graduating from highschool schoolhouse he joined the j1 conference order avispa fukuoka in 2000 although he debuted as a midfielder in 2001 he did not gambling much and the order was relegated to the j2 conference at the oddment of the 2001 season in 2002 he moved to the j2 order oita trinita he became a fixture histrion as a defensive midfielder and the order won the backup in 2002 and was promoted in 2003 he played many catch until 2005 in sep 2005 he moved to the j2 order montedio yamagata in 2006 he moved to the j2 order vissel kobe although he became a fixture histrion as a defensive midfielder his gradually was played less during the summertime in 2007 he moved to the nippon football conference order rosso kumamoto later roasso kumamoto based in his local area he played as a fixture histrion and the order was promoted to j2 in 2008 although he did not gambling as much he distillery played in many catch in 2010 he moved to indonesia and joined persela lamongan in july 2010 he returned to nippon and joined the j2 order giravanz kitakyushu he played often as a defensive midfielder and gist rear until 2012 when he retired
\end{enumerate}

\begin{enumerate}
\item If i have more pieces at the time of stalemate have i won
\item If single have more pieces at the time of stalemate have i won
\item If i have more opus at the time of stalemate have i won
\item If i have more opus at the sentence of deadlock have i won
\end{enumerate}

\begin{enumerate}
\item Stalemate is a situation in chess where the player whose turn it is to move is not in check and has no legal move stalemate results in a draw during the endgame stalemate is a resource that can enable the player with the inferior position to draw the game rather than lose in more complex positions stalemate is much rarer usually taking the form of a swindle that succeeds only if the superior side is inattentivecitation needed stalemate is also a common theme in endgame studies and other chess problemsthe outcome of a stalemate was standardized as a draw in the 19th century before this standardization its treatment varied widely including being deemed a win for the stalemating player a halfwin for that player or a loss for that player not being permitted and resulting in the stalemated player missing a turn stalemate rules vary in other games of the chess family
\item Deadlock is a situation in chess where the player whose turn it is to move is not in check and has no legal move stalemate results in a draw during the endgame stalemate is a resource that can enable the player with the inferior position to draw the game rather than lose in more complex positions stalemate is much rarer usually taking the form of a swindle that succeeds only if the superior side is inattentivecitation needed stalemate is also a common theme in endgame studies and other chess problemsthe outcome of a stalemate was standardized as a draw in the 19th century before this standardization its treatment varied widely including being deemed a win for the stalemating player a halfwin for that player or a loss for that player not being permitted and resulting in the stalemated player missing a turn stalemate rules vary in other games of the chess family
\item Deadlock is a situation in chess where the player whose turn it is to move is not in check and has no legal move stalemate results in a draw during the endgame stalemate is a resource that can enable the player with the inferior position to draw the game rather than lose in more complex positions stalemate is much rarer usually taking the form of a swindle that succeeds only if the superior side is inattentivecitation needed stalemate is also a common theme in endgame studies and other chess problemsthe outcome of a stalemate was standardized as a draw in the 19th century before this standardization its treatment varied widely including being deemed a win for the stalemating player a halfwin for that player or a loss for that player not being permitted and resulting in the stalemated player missing a turn stalemate rules vary in other games of the chess family
\item Deadlock is a billet in chess where the histrion whose round it is to motility is not in curb and has no legal motility deadlock effect in a attracter during the endgame deadlock is a imagination that can enable the histrion with the subscript stance to attracter the biz rather than lose in more composite stance deadlock is much rarer usually taking the flesh of a rig that succeeds only if the victor incline is inattentivecitation needed deadlock is also a park composition in endgame sketch and other chess problemsthe effect of a deadlock was standardized as a attracter in the 19th 100 before this normalization its handling varied widely including being deemed a win for the stalemating histrion a halfwin for that histrion or a expiration for that histrion not being permitted and resulting in the stalemated histrion missing a round deadlock regulation vary in other biz of the chess menage
\end{enumerate}

\begin{enumerate}
\item No stalemate is a drawn position it doesn't matter who has captured more pieces or is in a winning position
\item Nobelium stalemate is a drawn position it doesn't matter who has captured more pieces or is in a winning position
\item No deadlock is a drawn position it doesn't matter who has captured more pieces or is in a winning position
\item No deadlock is a drawn stance it doesn't topic who has captured more opus or is in a winning stance
\end{enumerate}

\begin{enumerate}
\item Given a reference text about lollapalooza where does it take place who started it and what is it
\item Precondition a reference text about lollapalooza where does it take place who started it and what is it
\item Precondition a reference text about lollapalooza where does it take place who started it and what is it
\item Precondition a character schoolbook about lollapalooza where does it payoff office who started it and what is it
\end{enumerate}

\begin{enumerate}
\item Lollapalooza llpluz lolla is an annual american fourday music festival held in grant park in chicago it originally started as a touring event in 1991 but several years later chicago became its permanent location music genres include but are not limited to alternative rock heavy metal punk rock hip hop and electronic dance music lollapalooza has also featured visual arts nonprofit organizations and political organizations the festival held in grant park hosts an estimated 400000 people each july and sells out annually lollapalooza is one of the largest and most iconic music festivals in the world and one of the longestrunning in the united stateslollapalooza was conceived and created in 1991 as a farewell tour by perry farrell singer of the group jane's addiction
\item Lollapalooza llpluz lolla is an yearbook american fourday music festival held in grant park in chicago it originally started as a touring event in 1991 but several years later chicago became its permanent location music genres include but are not limited to alternative rock heavy metal punk rock hip hop and electronic dance music lollapalooza has also featured visual arts nonprofit organizations and political organizations the festival held in grant park hosts an estimated 400000 people each july and sells out annually lollapalooza is one of the largest and most iconic music festivals in the world and one of the longestrunning in the united stateslollapalooza was conceived and created in 1991 as a farewell tour by perry farrell singer of the group jane's addiction
\item Lollapalooza llpluz lolla is an yearbook american fourday music festival held in grant park in chicago it originally started as a touring event in 1991 but several years later chicago became its permanent location music genres include but are not limited to alternative rock heavy metal punk rock hip hop and electronic dance music lollapalooza has also featured visual arts nonprofit organizations and political organizations the festival held in grant park hosts an estimated 400000 people each july and sells out annually lollapalooza is one of the largest and most iconic music festivals in the world and one of the longestrunning in the united stateslollapalooza was conceived and created in 1991 as a farewell tour by perry farrell singer of the group jane's addiction
\item Lollapalooza llpluz lolla is an yearbook american fourday medicine fete held in subsidisation parkland in boodle it originally started as a touring effect in 1991 but several class later boodle became its perm emplacement medicine genres include but are not limited to option rock'n'roll heavy alloy kindling rock'n'roll coxa hop and electronic saltation medicine lollapalooza has also featured visual prowess nonprofit formation and political formation the fete held in subsidisation parkland server an estimated 400000 citizenry each july and sells out annually lollapalooza is 1 of the largest and most iconic medicine festival in the humanity and 1 of the longestrunning in the united stateslollapalooza was conceived and created in 1991 as a parting spell by perry farrell vocalist of the radical jane's dependency
\end{enumerate}

\begin{enumerate}
\item Lollapalooze is an annual musical festival held in grant park in chicago illinois it was started in 1991 as a farewell tour by perry farrell singe of the group jane's addiction the festival includes an array of musical genres including alternative rock heavy metal punk rock hip hop and electronic dance music the festivals welcomes an estimated 400000 people each year and sells out annually some notable headliners include the red hot chili peppers chance the rapper metallica and lady gage lollapalooza is one of the largest and most iconic festivals in the world and a staple of chicago
\item Lollapalooze is an yearbook musical festival held in grant park in chicago illinois it was started in 1991 as a farewell tour by perry farrell singe of the group jane's addiction the festival includes an array of musical genres including alternative rock heavy metal punk rock hip hop and electronic dance music the festivals welcomes an estimated 400000 people each year and sells out annually some notable headliners include the red hot chili peppers chance the rapper metallica and lady gage lollapalooza is one of the largest and most iconic festivals in the world and a staple of chicago
\item Lollapalooze is an yearbook musical festival held in grant park in chicago illinois it was started in 1991 as a farewell tour by perry farrell singe of the group jane's addiction the festival includes an array of musical genres including alternative rock heavy metal punk rock hip hop and electronic dance music the festivals welcomes an estimated 400000 people each year and sells out annually some notable headliners include the red hot chili peppers chance the rapper metallica and lady gage lollapalooza is one of the largest and most iconic festivals in the world and a staple of chicago
\item Lollapalooze is an yearbook musical fete held in subsidisation parkland in boodle illinois it was started in 1991 as a parting spell by perry farrell scorch of the radical jane's dependency the fete includes an regalia of musical genres including option rock'n'roll heavy alloy kindling rock'n'roll coxa hop and electronic saltation medicine the festival welcomes an estimated 400000 citizenry each yr and sells out annually some luminary headliner include the bolshy hot chile capsicum hazard the knocker metallica and dame dope lollapalooza is 1 of the largest and most iconic festival in the humanity and a basic of boodle
\end{enumerate}

\begin{enumerate}
\item Why mobile is bad for human
\item Wherefore mobile is bad for human
\item Why mobile is bad for homo
\item Why mobile is bad for homo
\end{enumerate}

\begin{enumerate}
\item We are always engaged one phone which is not good
\item We ar always engaged one phone which is not good
\item We are always engaged 1 phone which is not good
\item We are always engaged 1 telephone which is not goodness
\end{enumerate}

\begin{enumerate}
\item Who was john moses browning
\item Who washington john moses browning
\item Who was lav moses browning
\item Who was lav moses toasting
\end{enumerate}

\begin{enumerate}
\item John moses browning january 23 1855 november 26 1926 was an american firearm designer who developed many varieties of military and civilian firearms cartridges and gun mechanisms many of which are still in use around the world he made his first firearm at age 13 in his father's gun shop and was awarded the first of his 128 firearm patents on october 7 1879 at the age of 24 he is regarded as one of the most successful firearms designers of the 19th and 20th centuries and pioneered the development of modern repeating semiautomatic and automatic firearmsbrowning influenced nearly all categories of firearms design especially the autoloading of ammunition he invented or made significant improvements to singleshot leveraction and pumpaction rifles and shotguns he developed the first reliable and compact autoloading pistols by inventing the telescoping bolt then integrating the bolt and barrel shroud into what is known as the pistol slide browning's telescoping bolt design is now found on nearly every modern semiautomatic pistol as well as several modern fully automatic weapons he also developed the first gasoperated firearm the coltbrowning model 1895 machine gun a system that surpassed mechanical recoil operation to become the standard for most highpower selfloading firearm designs worldwide he also made significant contributions to automatic cannon developmentbrowning's most successful designs include the m1911 pistol the watercooled m1917 the aircooled m1919 and heavy m2 machine guns the m1918 browning automatic rifle and the browning auto5 the first semiautomatic shotgun some of these arms are still manufactured often with only minor changes in detail and cosmetics to those assembled by browning or his licensees the browningdesigned m1911 and hipower are some of the most copied firearms in the world
\item Lav moses browning january 23 1855 november 26 1926 was an american firearm designer who developed many varieties of military and civilian firearms cartridges and gun mechanisms many of which are still in use around the world he made his first firearm at age 13 in his father's gun shop and was awarded the first of his 128 firearm patents on october 7 1879 at the age of 24 he is regarded as one of the most successful firearms designers of the 19th and 20th centuries and pioneered the development of modern repeating semiautomatic and automatic firearmsbrowning influenced nearly all categories of firearms design especially the autoloading of ammunition he invented or made significant improvements to singleshot leveraction and pumpaction rifles and shotguns he developed the first reliable and compact autoloading pistols by inventing the telescoping bolt then integrating the bolt and barrel shroud into what is known as the pistol slide browning's telescoping bolt design is now found on nearly every modern semiautomatic pistol as well as several modern fully automatic weapons he also developed the first gasoperated firearm the coltbrowning model 1895 machine gun a system that surpassed mechanical recoil operation to become the standard for most highpower selfloading firearm designs worldwide he also made significant contributions to automatic cannon developmentbrowning's most successful designs include the m1911 pistol the watercooled m1917 the aircooled m1919 and heavy m2 machine guns the m1918 browning automatic rifle and the browning auto5 the first semiautomatic shotgun some of these arms are still manufactured often with only minor changes in detail and cosmetics to those assembled by browning or his licensees the browningdesigned m1911 and hipower are some of the most copied firearms in the world
\item Lav moses browning january 23 1855 november 26 1926 was an american firearm designer who developed many varieties of military and civilian firearms cartridges and gun mechanisms many of which are still in use around the world he made his first firearm at age 13 in his father's gun shop and was awarded the first of his 128 firearm patents on october 7 1879 at the age of 24 he is regarded as one of the most successful firearms designers of the 19th and 20th centuries and pioneered the development of modern repeating semiautomatic and automatic firearmsbrowning influenced nearly all categories of firearms design especially the autoloading of ammunition he invented or made significant improvements to singleshot leveraction and pumpaction rifles and shotguns he developed the first reliable and compact autoloading pistols by inventing the telescoping bolt then integrating the bolt and barrel shroud into what is known as the pistol slide browning's telescoping bolt design is now found on nearly every modern semiautomatic pistol as well as several modern fully automatic weapons he also developed the first gasoperated firearm the coltbrowning model 1895 machine gun a system that surpassed mechanical recoil operation to become the standard for most highpower selfloading firearm designs worldwide he also made significant contributions to automatic cannon developmentbrowning's most successful designs include the m1911 pistol the watercooled m1917 the aircooled m1919 and heavy m2 machine guns the m1918 browning automatic rifle and the browning auto5 the first semiautomatic shotgun some of these arms are still manufactured often with only minor changes in detail and cosmetics to those assembled by browning or his licensees the browningdesigned m1911 and hipower are some of the most copied firearms in the world
\item Lav moses toasting jan twenty-three 1855 nov xxvi 1926 was an american piece architect who developed many mixture of military and civilian piece cartridge and shooter mechanisms many of which are distillery in utilization around the humanity he made his get-go piece at eld thirteen in his father's shooter workshop and was awarded the get-go of his 128 piece patents on oct heptad 1879 at the eld of xxiv he is regarded as 1 of the most successful piece architect of the 19th and 20th c and pioneered the developing of bodoni repetition semiautomatic and automatic firearmsbrowning influenced nearly all class of piece intent especially the autoloading of ammo he invented or made significant melioration to singleshot leveraction and pumpaction rifles and shotgun he developed the get-go reliable and covenant autoloading pistol by inventing the telescoping deadbolt then desegregation the deadbolt and drum winding-sheet into what is known as the handgun chute browning's telescoping deadbolt intent is now found on nearly every bodoni semiautomatic handgun as wellspring as several bodoni fully automatic artillery he also developed the get-go gasoperated piece the coltbrowning simulation 1895 motorcar shooter a scheme that surpassed mechanical kick procedure to become the touchstone for most highpower selfloading piece plan worldwide he also made significant contribution to automatic shank developmentbrowning's most successful plan include the m1911 handgun the watercooled m1917 the aircooled m1919 and heavy m2 motorcar triggerman the m1918 toasting automatic rifle and the toasting auto5 the get-go semiautomatic scattergun some of these branch are distillery manufactured often with only youngster variety in item and cosmetics to those assembled by toasting or his licensee the browningdesigned m1911 and hipower are some of the most copied piece in the humanity
\end{enumerate}

\begin{enumerate}
\item John moses browning is one of the most wellknown designer of modern firearms he started building firearms in his father's shop at the age of 13 and was awarded his first patent when he was 24he designed the first reliable automatic pistol and the first gasoperated firearm as well inventing or improving singleshot leveraction and pumpaction rifles and shotgunstoday he is most wellknown for the m1911 pistol the browning automatic rifle and the auto5 shotgun all of which are in still in current production in either their original design or with minor changes his m1911 and hipower pistols designs are some of the most reproduced firearms in the world today
\item Lav moses browning is one of the most wellknown designer of modern firearms he started building firearms in his father's shop at the age of 13 and was awarded his first patent when he was 24he designed the first reliable automatic pistol and the first gasoperated firearm as well inventing or improving singleshot leveraction and pumpaction rifles and shotgunstoday he is most wellknown for the m1911 pistol the browning automatic rifle and the auto5 shotgun all of which are in still in current production in either their original design or with minor changes his m1911 and hipower pistols designs are some of the most reproduced firearms in the world today
\item Lav moses browning is one of the most wellknown designer of modern firearms he started building firearms in his father's shop at the age of 13 and was awarded his first patent when he was 24he designed the first reliable automatic pistol and the first gasoperated firearm as well inventing or improving singleshot leveraction and pumpaction rifles and shotgunstoday he is most wellknown for the m1911 pistol the browning automatic rifle and the auto5 shotgun all of which are in still in current production in either their original design or with minor changes his m1911 and hipower pistols designs are some of the most reproduced firearms in the world today
\item Lav moses toasting is 1 of the most wellknown architect of bodoni piece he started edifice piece in his father's workshop at the eld of thirteen and was awarded his get-go patent when he was 24he designed the get-go reliable automatic handgun and the get-go gasoperated piece as wellspring inventing or improving singleshot leveraction and pumpaction rifles and shotgunstoday he is most wellknown for the m1911 handgun the toasting automatic rifle and the auto5 scattergun all of which are in distillery in flow product in either their archetype intent or with youngster variety his m1911 and hipower pistol plan are some of the most reproduced piece in the humanity today
\end{enumerate}

\begin{enumerate}
\item Thomas jefferson april 13 1743 july 4 1826 was an american statesman diplomat lawyer architect philosopher and founding father who served as the third president of the united states from 1801 to 1809 among the committee of five charged by the second continental congress with authoring the declaration of independence jefferson was the declaration's primary author following the american revolutionary war and prior to becoming the nation's third president in 1801 jefferson was the first united states secretary of state under george washington and then the nation's second vice president under john adams
\item Thomas jefferson apr 13 1743 july 4 1826 was an american statesman diplomat lawyer architect philosopher and founding father who served as the third president of the united states from 1801 to 1809 among the committee of five charged by the second continental congress with authoring the declaration of independence jefferson was the declaration's primary author following the american revolutionary war and prior to becoming the nation's third president in 1801 jefferson was the first united states secretary of state under george washington and then the nation's second vice president under john adams
\item Thomas jefferson apr 13 1743 july 4 1826 was an american statesman diplomat lawyer architect philosopher and founding father who served as the third president of the united states from 1801 to 1809 among the committee of five charged by the second continental congress with authoring the declaration of independence jefferson was the declaration's primary author following the american revolutionary war and prior to becoming the nation's third president in 1801 jefferson was the first united states secretary of state under george washington and then the nation's second vice president under john adams
\item Thomas jefferson apr thirteen 1743 july quaternity 1826 was an american solon diplomatist attorney designer philosopher and institution forefather who served as the tierce chairman of the united commonwealth from 1801 to 1809 among the commission of fin charged by the secondment continental coitus with authoring the proclamation of independency jefferson was the declaration's primary writer chase the american subversive warfare and prior to becoming the nation's tierce chairman in 1801 jefferson was the get-go united commonwealth secretaire of commonwealth under george capital and then the nation's secondment frailty chairman under lav go
\end{enumerate}

\begin{enumerate}
\item Thomas jefferson april 13 1743 july 4 1826 was an american statesman diplomat lawyer architect philosopher and founding father who served as the third president of the united states from 1801 to 1809 among the committee of five charged by the second continental congress with authoring the declaration of independence jefferson was the declaration's primary author following the american revolutionary war and prior to becoming the nation's third president in 1801 jefferson was the first united states secretary of state under george washington and then the nation's second vice president under john adamsamong the nation's founding fathers jefferson is considered unmatched in his intellectual depth and breadth his passionate writings and advocacy for human rights including freedom of thought speech and religion were a leading inspiration behind the american revolution which ultimately gave rise to the american revolutionary war american independence and the united states constitution jefferson's ideas were globally influential in shaping and inspiring the age of enlightenment which proved transformational in the late 17th and 18th centuries he was a leading proponent of democracy republicanism and individual rights and produced formative documents and decisions at the state national and international levelsduring the american revolution jefferson represented virginia in the second continental congress in philadelphia which adopted the declaration of independence on july 4 1776 as a virginia legislator he drafted a state law for religious freedom he served as the second governor of virginia from 1779 to 1781 during the revolutionary war in 1785 jefferson was appointed the united states minister to france and subsequently the nation's first secretary of state under president george washington from 1790 to 1793 jefferson and james madison organized the democraticrepublican party to oppose the federalist party during the formation of the first party system with madison he anonymously wrote the kentucky and virginia resolutions in 1798 and 1799 which sought to strengthen states' rights by nullifying the federal alien and sedition actsjefferson and federalist john adams became friends as well as political rivals serving in the continental congress and drafting the declaration of independence together in the 1796 presidential election between the two jefferson came in second which according to electoral procedure at the time made him vice president to adams jefferson challenged adams again in 1800 and won the presidency after his term in office jefferson eventually reconciled with adams and they shared a correspondence that lasted 14 years he and adams both died on the same day july 4 1826 which was also the 50th anniversary of declaration of independenceas president jefferson pursued the nation's shipping and trade interests against barbary pirates and aggressive british trade policies starting in 1803 he promoted a western expansionist policy with the louisiana purchase which doubled the nation's claimed land area to make room for settlement jefferson began the process of indian tribal removal from the newly acquired territory as a result of peace negotiations with france his administration reduced military forces he was reelected in 1804 but his second term was beset with difficulties at home including the trial of former vice president aaron burr in 1807 american foreign trade was diminished when jefferson implemented the embargo act in response to british threats to us shipping the same year jefferson signed the act prohibiting importation of slavesjefferson was a plantation owner lawyer and politician and mastered many disciplines including surveying mathematics horticulture and mechanics he was also an architect in the palladian tradition jefferson's keen interest in religion and philosophy led to his appointment as president of the american philosophical society he largely shunned organized religion but was influenced by christianity epicureanism and deism jefferson rejected fundamental christianity denying christ's divinity a philologist jefferson knew several languages he was a prolific letter writer and corresponded with many prominent people including edward carrington john taylor of caroline and james madison in 1785 jefferson authored notes on the state of virginia considered perhaps the most important american book published before 1800 jefferson championed the ideals values and teachings of the enlightenmentsince the 1790s jefferson was rumored to have had children by his sisterinlaw and slave sally hemings leading to what is known as the jeffersonhemings controversy a 1998 dna test concluded that one of sally hemings's children eston hemings was of the jefferson male line according to scholarly consensus based on documentary and statistical evaluation as well as oral history jefferson probably fathered at least six children with hemings including four that survived to adulthoodafter retiring from public office jefferson founded the university of virginia presidential scholars and historians generally praise jefferson's public achievements including his advocacy of religious freedom and tolerance in virginia his peaceful acquisition of the louisiana territory from france without war or controversy and his ambitious and successful lewis and clark expedition some modern historians are critical of jefferson's personal involvement with slavery jefferson is consistently ranked among the top ten presidents of american history
\item Thomas jefferson apr 13 1743 july 4 1826 was an american statesman diplomat lawyer architect philosopher and founding father who served as the third president of the united states from 1801 to 1809 among the committee of five charged by the second continental congress with authoring the declaration of independence jefferson was the declaration's primary author following the american revolutionary war and prior to becoming the nation's third president in 1801 jefferson was the first united states secretary of state under george washington and then the nation's second vice president under john adamsamong the nation's founding fathers jefferson is considered unmatched in his intellectual depth and breadth his passionate writings and advocacy for human rights including freedom of thought speech and religion were a leading inspiration behind the american revolution which ultimately gave rise to the american revolutionary war american independence and the united states constitution jefferson's ideas were globally influential in shaping and inspiring the age of enlightenment which proved transformational in the late 17th and 18th centuries he was a leading proponent of democracy republicanism and individual rights and produced formative documents and decisions at the state national and international levelsduring the american revolution jefferson represented virginia in the second continental congress in philadelphia which adopted the declaration of independence on july 4 1776 as a virginia legislator he drafted a state law for religious freedom he served as the second governor of virginia from 1779 to 1781 during the revolutionary war in 1785 jefferson was appointed the united states minister to france and subsequently the nation's first secretary of state under president george washington from 1790 to 1793 jefferson and james madison organized the democraticrepublican party to oppose the federalist party during the formation of the first party system with madison he anonymously wrote the kentucky and virginia resolutions in 1798 and 1799 which sought to strengthen states' rights by nullifying the federal alien and sedition actsjefferson and federalist john adams became friends as well as political rivals serving in the continental congress and drafting the declaration of independence together in the 1796 presidential election between the two jefferson came in second which according to electoral procedure at the time made him vice president to adams jefferson challenged adams again in 1800 and won the presidency after his term in office jefferson eventually reconciled with adams and they shared a correspondence that lasted 14 years he and adams both died on the same day july 4 1826 which was also the 50th anniversary of declaration of independenceas president jefferson pursued the nation's shipping and trade interests against barbary pirates and aggressive british trade policies starting in 1803 he promoted a western expansionist policy with the louisiana purchase which doubled the nation's claimed land area to make room for settlement jefferson began the process of indian tribal removal from the newly acquired territory as a result of peace negotiations with france his administration reduced military forces he was reelected in 1804 but his second term was beset with difficulties at home including the trial of former vice president aaron burr in 1807 american foreign trade was diminished when jefferson implemented the embargo act in response to british threats to us shipping the same year jefferson signed the act prohibiting importation of slavesjefferson was a plantation owner lawyer and politician and mastered many disciplines including surveying mathematics horticulture and mechanics he was also an architect in the palladian tradition jefferson's keen interest in religion and philosophy led to his appointment as president of the american philosophical society he largely shunned organized religion but was influenced by christianity epicureanism and deism jefferson rejected fundamental christianity denying christ's divinity a philologist jefferson knew several languages he was a prolific letter writer and corresponded with many prominent people including edward carrington john taylor of caroline and james madison in 1785 jefferson authored notes on the state of virginia considered perhaps the most important american book published before 1800 jefferson championed the ideals values and teachings of the enlightenmentsince the 1790s jefferson was rumored to have had children by his sisterinlaw and slave sally hemings leading to what is known as the jeffersonhemings controversy a 1998 dna test concluded that one of sally hemings's children eston hemings was of the jefferson male line according to scholarly consensus based on documentary and statistical evaluation as well as oral history jefferson probably fathered at least six children with hemings including four that survived to adulthoodafter retiring from public office jefferson founded the university of virginia presidential scholars and historians generally praise jefferson's public achievements including his advocacy of religious freedom and tolerance in virginia his peaceful acquisition of the louisiana territory from france without war or controversy and his ambitious and successful lewis and clark expedition some modern historians are critical of jefferson's personal involvement with slavery jefferson is consistently ranked among the top ten presidents of american history
\item Thomas jefferson apr 13 1743 july 4 1826 was an american statesman diplomat lawyer architect philosopher and founding father who served as the third president of the united states from 1801 to 1809 among the committee of five charged by the second continental congress with authoring the declaration of independence jefferson was the declaration's primary author following the american revolutionary war and prior to becoming the nation's third president in 1801 jefferson was the first united states secretary of state under george washington and then the nation's second vice president under john adamsamong the nation's founding fathers jefferson is considered unmatched in his intellectual depth and breadth his passionate writings and advocacy for human rights including freedom of thought speech and religion were a leading inspiration behind the american revolution which ultimately gave rise to the american revolutionary war american independence and the united states constitution jefferson's ideas were globally influential in shaping and inspiring the age of enlightenment which proved transformational in the late 17th and 18th centuries he was a leading proponent of democracy republicanism and individual rights and produced formative documents and decisions at the state national and international levelsduring the american revolution jefferson represented virginia in the second continental congress in philadelphia which adopted the declaration of independence on july 4 1776 as a virginia legislator he drafted a state law for religious freedom he served as the second governor of virginia from 1779 to 1781 during the revolutionary war in 1785 jefferson was appointed the united states minister to france and subsequently the nation's first secretary of state under president george washington from 1790 to 1793 jefferson and james madison organized the democraticrepublican party to oppose the federalist party during the formation of the first party system with madison he anonymously wrote the kentucky and virginia resolutions in 1798 and 1799 which sought to strengthen states' rights by nullifying the federal alien and sedition actsjefferson and federalist john adams became friends as well as political rivals serving in the continental congress and drafting the declaration of independence together in the 1796 presidential election between the two jefferson came in second which according to electoral procedure at the time made him vice president to adams jefferson challenged adams again in 1800 and won the presidency after his term in office jefferson eventually reconciled with adams and they shared a correspondence that lasted 14 years he and adams both died on the same day july 4 1826 which was also the 50th anniversary of declaration of independenceas president jefferson pursued the nation's shipping and trade interests against barbary pirates and aggressive british trade policies starting in 1803 he promoted a western expansionist policy with the louisiana purchase which doubled the nation's claimed land area to make room for settlement jefferson began the process of indian tribal removal from the newly acquired territory as a result of peace negotiations with france his administration reduced military forces he was reelected in 1804 but his second term was beset with difficulties at home including the trial of former vice president aaron burr in 1807 american foreign trade was diminished when jefferson implemented the embargo act in response to british threats to us shipping the same year jefferson signed the act prohibiting importation of slavesjefferson was a plantation owner lawyer and politician and mastered many disciplines including surveying mathematics horticulture and mechanics he was also an architect in the palladian tradition jefferson's keen interest in religion and philosophy led to his appointment as president of the american philosophical society he largely shunned organized religion but was influenced by christianity epicureanism and deism jefferson rejected fundamental christianity denying christ's divinity a philologist jefferson knew several languages he was a prolific letter writer and corresponded with many prominent people including edward carrington john taylor of caroline and james madison in 1785 jefferson authored notes on the state of virginia considered perhaps the most important american book published before 1800 jefferson championed the ideals values and teachings of the enlightenmentsince the 1790s jefferson was rumored to have had children by his sisterinlaw and slave sally hemings leading to what is known as the jeffersonhemings controversy a 1998 dna test concluded that one of sally hemings's children eston hemings was of the jefferson male line according to scholarly consensus based on documentary and statistical evaluation as well as oral history jefferson probably fathered at least six children with hemings including four that survived to adulthoodafter retiring from public office jefferson founded the university of virginia presidential scholars and historians generally praise jefferson's public achievements including his advocacy of religious freedom and tolerance in virginia his peaceful acquisition of the louisiana territory from france without war or controversy and his ambitious and successful lewis and clark expedition some modern historians are critical of jefferson's personal involvement with slavery jefferson is consistently ranked among the top ten presidents of american history
\item Thomas jefferson apr thirteen 1743 july quaternity 1826 was an american solon diplomatist attorney designer philosopher and institution forefather who served as the tierce chairman of the united commonwealth from 1801 to 1809 among the commission of fin charged by the secondment continental coitus with authoring the proclamation of independency jefferson was the declaration's primary writer chase the american subversive warfare and prior to becoming the nation's tierce chairman in 1801 jefferson was the get-go united commonwealth secretaire of commonwealth under george capital and then the nation's secondment frailty chairman under lav adamsamong the nation's institution fatherhood jefferson is considered unmatched in his intellect profundity and largeness his passionate writing and protagonism for homo right including exemption of persuasion delivery and faith were a lead stirring backside the american gyration which ultimately gave raise to the american subversive warfare american independency and the united commonwealth composition jefferson's estimation were globally influential in formation and inspiring the eld of nirvana which proved transformational in the late 17th and 18th c he was a lead exponent of commonwealth republicanism and person right and produced formative documents and decisiveness at the commonwealth subject and international levelsduring the american gyration jefferson represented va in the secondment continental coitus in philadelphia which adopted the proclamation of independency on july quaternity 1776 as a va legislator he drafted a commonwealth constabulary for religious exemption he served as the secondment regulator of va from 1779 to 1781 during the subversive warfare in 1785 jefferson was appointed the united commonwealth curate to france and subsequently the nation's get-go secretaire of commonwealth under chairman george capital from 1790 to 1793 jefferson and james madison organized the democraticrepublican company to oppose the federalist company during the establishment of the get-go company scheme with madison he anonymously wrote the ky and va closure in 1798 and 1799 which sought to strengthen states' right by nullifying the fed unknown and sedition actsjefferson and federalist lav go became supporter as wellspring as political competition portion in the continental coitus and drawing the proclamation of independency together in the 1796 presidential election between the 2 jefferson came in secondment which according to electoral subroutine at the sentence made him frailty chairman to go jefferson challenged go again in 1800 and won the administration after his condition in post jefferson eventually reconciled with go and they shared a symmetry that lasted xiv class he and go both died on the same daylight july quaternity 1826 which was also the 50th anniversary of proclamation of independenceas chairman jefferson chased the nation's transportation and barter pursuit against barbary pirate and aggressive british barter policies start in 1803 he promoted a western expansionist insurance with the la leverage which doubled the nation's claimed kingdom arena to shuffling way for closure jefferson began the summons of amerind tribal remotion from the newly acquired soil as a effect of pacification negotiation with france his government reduced military forcefulness he was reelected in 1804 but his secondment condition was beset with difficulty at menage including the visitation of former frailty chairman aaron bur in 1807 american foreign barter was diminished when jefferson implemented the embargo turn in reception to british menace to uracil transportation the same yr jefferson signed the turn prohibiting import of slavesjefferson was a woodlet possessor attorney and politico and mastered many correction including surveying maths gardening and mechanic he was also an designer in the palladian custom jefferson's keen pastime in faith and doctrine led to his engagement as chairman of the american philosophical fellowship he largely shunned organized faith but was influenced by christendom epicureanism and deism jefferson rejected fundamental christendom denying christ's theology a philologue jefferson knew several lyric he was a prolific missive author and corresponded with many prominent citizenry including edward carrington lav taylor of caroline and james madison in 1785 jefferson authored note on the commonwealth of va considered perhaps the most important american leger published before 1800 jefferson championed the saint value and instruction of the enlightenmentsince the 1790s jefferson was rumored to have had youngster by his sisterinlaw and striver quip hemings lead to what is known as the jeffersonhemings contention a 1998 dna tryout concluded that 1 of quip hemings's youngster eston hemings was of the jefferson male contrast according to scholarly consensus based on docudrama and statistical rating as wellspring as viva story jefferson probably fathered at least vi youngster with hemings including quaternity that survived to adulthoodafter retiring from world post jefferson founded the university of va presidential student and historian generally extolment jefferson's world achievement including his protagonism of religious exemption and margin in va his peaceful acquirement of the la soil from france without warfare or contention and his ambitious and successful lewis and clark outing some bodoni historian are critical of jefferson's personal engagement with bondage jefferson is consistently ranked among the upside x chair of american story
\end{enumerate}

\begin{enumerate}
\item Who was kyle van zyl playing against when he scored 36 of hisa teams 61 points
\item Who washington kyle van zyl playing against when he scored 36 of hisa teams 61 points
\item Who was kyle caravan zyl playing against when he scored 36 of hisa teams 61 points
\item Who was kyle caravan zyl performing against when he scored 36 of hisa squad 61 period
\end{enumerate}

\begin{enumerate}
\item Van zyl joined the eastern province kings academy where he played for the eastern province u19 side in the 2010 under19 provincial championship he was a key player for the eastern province u21 side in the 2012 under21 provincial championship scoring 71 points in eight appearances van zyl was under the top saru performers scoring the most tries at 6 in the 2012 provincial under 21 in the rugby junior provincialsthis included a record and a remarkable personal haul in their opening match when he scored 36 of his team's points in a 613 victory over boland u21 consisting of four tries and eight conversions and was awarded man of the match
\item Caravan zyl joined the eastern province kings academy where he played for the eastern province u19 side in the 2010 under19 provincial championship he was a key player for the eastern province u21 side in the 2012 under21 provincial championship scoring 71 points in eight appearances van zyl was under the top saru performers scoring the most tries at 6 in the 2012 provincial under 21 in the rugby junior provincialsthis included a record and a remarkable personal haul in their opening match when he scored 36 of his team's points in a 613 victory over boland u21 consisting of four tries and eight conversions and was awarded man of the match
\item Caravan zyl joined the eastern province kings academy where he played for the eastern province u19 side in the 2010 under19 provincial championship he was a key player for the eastern province u21 side in the 2012 under21 provincial championship scoring 71 points in eight appearances van zyl was under the top saru performers scoring the most tries at 6 in the 2012 provincial under 21 in the rugby junior provincialsthis included a record and a remarkable personal haul in their opening match when he scored 36 of his team's points in a 613 victory over boland u21 consisting of four tries and eight conversions and was awarded man of the match
\item Caravan zyl joined the eastern responsibility mogul academy where he played for the eastern responsibility u19 incline in the 2010 under19 peasant backup he was a samara histrion for the eastern responsibility u21 incline in the 2012 under21 peasant backup marking 71 period in 8 show caravan zyl was under the upside saru performers marking the most attempt at vi in the 2012 peasant under twenty-one in the rugger jr provincialsthis included a book and a remarkable personal catch in their initiative catch when he scored 36 of his team's period in a 613 triumph over boland u21 consisting of quaternity attempt and 8 conversion and was awarded humanity of the catch
\end{enumerate}

\begin{enumerate}
\item Kyle van zyl was playing against boland u21 when he scored 36 points leading his team to victory in a 613 win
\item Kyle caravan zyl was playing against boland u21 when he scored 36 points leading his team to victory in a 613 win
\item Kyle caravan zyl was playing against boland u21 when he scored 36 points leading his team to victory in a 613 win
\item Kyle caravan zyl was performing against boland u21 when he scored 36 period lead his squad to triumph in a 613 win
\end{enumerate}

\begin{enumerate}
\item From the passage list down the areas for which dar es salaam is tanzania's most prominent city list the results in comma separated format
\item From the transit list down the areas for which dar es salaam is tanzania's most prominent city list the results in comma separated format
\item From the transit list down the areas for which dar es salaam is tanzania's most prominent city list the results in comma separated format
\item From the transit lean down the orbit for which dar einsteinium salaam is tanzania's most prominent city lean the effect in comma separated formatting
\end{enumerate}

\begin{enumerate}
\item Dar es salaam dr s slm from arabic romanized dr esselm lit 'abode of peace' or commonly known as dar is the largest city and financial hub of tanzania it is also the capital of dar es salaam region with a population of over six million people dar is the largest city in east africa and the seventhlargest in africa located on the swahili coast dar es salaam is an important economic centre and is one of the fastestgrowing cities in the worldthe town was founded by majid bin said the first sultan of zanzibar in 1865 or 1866 it was the main administrative and commercial center of german east africa tanganyika and tanzania the decision was made in 1974 to move the capital to dodoma and was officially completed in 1996dar es salaam is tanzania's most prominent city for arts fashion media film television and finance it is the capital of the coextensive dar es salaam region one of tanzania's 31 administrative regions and consists of five districts kinondoni in the north ilala in the centre ubungo and temeke in the south and kigamboni in the east across the kurasini estuary
\item Dar eastward salaam dr s slm from arabic romanized dr esselm lit 'abode of peace' or commonly known as dar is the largest city and financial hub of tanzania it is also the capital of dar es salaam region with a population of over six million people dar is the largest city in east africa and the seventhlargest in africa located on the swahili coast dar es salaam is an important economic centre and is one of the fastestgrowing cities in the worldthe town was founded by majid bin said the first sultan of zanzibar in 1865 or 1866 it was the main administrative and commercial center of german east africa tanganyika and tanzania the decision was made in 1974 to move the capital to dodoma and was officially completed in 1996dar es salaam is tanzania's most prominent city for arts fashion media film television and finance it is the capital of the coextensive dar es salaam region one of tanzania's 31 administrative regions and consists of five districts kinondoni in the north ilala in the centre ubungo and temeke in the south and kigamboni in the east across the kurasini estuary
\item Dar einsteinium salaam dr s slm from arabic romanized dr esselm lit 'abode of peace' or commonly known as dar is the largest city and financial hub of tanzania it is also the capital of dar es salaam region with a population of over six million people dar is the largest city in east africa and the seventhlargest in africa located on the swahili coast dar es salaam is an important economic centre and is one of the fastestgrowing cities in the worldthe town was founded by majid bin said the first sultan of zanzibar in 1865 or 1866 it was the main administrative and commercial center of german east africa tanganyika and tanzania the decision was made in 1974 to move the capital to dodoma and was officially completed in 1996dar es salaam is tanzania's most prominent city for arts fashion media film television and finance it is the capital of the coextensive dar es salaam region one of tanzania's 31 administrative regions and consists of five districts kinondoni in the north ilala in the centre ubungo and temeke in the south and kigamboni in the east across the kurasini estuary
\item Dar einsteinium salaam dr s slm from arabic romanized dr esselm literature 'abode of peace' or commonly known as dar is the largest city and financial hub of tanzania it is also the chapiter of dar einsteinium salaam area with a universe of over vi meg citizenry dar is the largest city in eastward africa and the seventhlargest in africa located on the swahili seashore dar einsteinium salaam is an important economic gist and is 1 of the fastestgrowing metropolis in the worldthe township was founded by majid binful said the get-go sultan of zanzibar in 1865 or 1866 it was the briny administrative and commercial gist of german eastward africa tanganyika and tanzania the determination was made in 1974 to motility the chapiter to dodoma and was officially completed in 1996dar einsteinium salaam is tanzania's most prominent city for prowess mode medium picture telly and finance it is the chapiter of the coextensive dar einsteinium salaam area 1 of tanzania's 31 administrative area and consists of fin dominion kinondoni in the union ilala in the gist ubungo and temeke in the southward and kigamboni in the eastward across the kurasini estuary
\end{enumerate}

\begin{enumerate}
\item Arts fashion media film television finance
\item Prowess fashion media film television finance
\item Prowess fashion media film television finance
\item Prowess mode medium picture telly finance
\end{enumerate}

\begin{enumerate}
\item A polygon is a form in geometry it is a single dimensional plane made of connecting lines and any number of vertices it is a closed chain of connected line segments or edges the vertices of the polygon are formed where two edges meet examples of polygons are hexagons pentagons and octagons any plane that does not contain edges or vertices is not a polygon an example of a nonpolygon is a circle
\item Angstrom polygon is a form in geometry it is a single dimensional plane made of connecting lines and any number of vertices it is a closed chain of connected line segments or edges the vertices of the polygon are formed where two edges meet examples of polygons are hexagons pentagons and octagons any plane that does not contain edges or vertices is not a polygon an example of a nonpolygon is a circle
\item A polygon is a flesh in geometry it is a single dimensional plane made of connecting lines and any number of vertices it is a closed chain of connected line segments or edges the vertices of the polygon are formed where two edges meet examples of polygons are hexagons pentagons and octagons any plane that does not contain edges or vertices is not a polygon an example of a nonpolygon is a circle
\item A polygon is a flesh in geometry it is a one dimensional planer made of connecting crinkle and any turn of acme it is a closed concatenation of connected contrast segment or sharpness the acme of the polygon are formed where 2 sharpness meet model of polygons are hexagons pentagons and octagons any planer that does not contain sharpness or acme is not a polygon an case of a nonpolygon is a round
\end{enumerate}

\begin{enumerate}
\item How do i start running
\item How brawl i start running
\item How do i get-go running
\item How do i get-go run
\end{enumerate}

\begin{enumerate}
\item Make sure you get comfortable running shoes and attire start with achievable goal in mind like a 5k race if you never ran before start gradually from a walk to brisk walk light jog aiming for 1530mins initially slowly increase your running time and distance as your fitness level improves one of the most important things is cool down and gentle stretching always listen to your body and take rest days when needed to prevent injury
\item Shuffling sure you get comfortable running shoes and attire start with achievable goal in mind like a 5k race if you never ran before start gradually from a walk to brisk walk light jog aiming for 1530mins initially slowly increase your running time and distance as your fitness level improves one of the most important things is cool down and gentle stretching always listen to your body and take rest days when needed to prevent injury
\item Shuffling sure you get comfortable running shoes and attire start with achievable goal in mind like a 5k race if you never ran before start gradually from a walk to brisk walk light jog aiming for 1530mins initially slowly increase your running time and distance as your fitness level improves one of the most important things is cool down and gentle stretching always listen to your body and take rest days when needed to prevent injury
\item Shuffling sure you get comfortable run shoe and garb get-go with achievable destination in psyche ilk a 5k airstream if you never ran before get-go gradually from a walkway to brisk walkway lighting lope aiming for 1530mins initially slowly step-up your run sentence and space as your fittingness grade improves 1 of the most important affair is sang-froid down and gentle stretch always listen to your torso and payoff residual day when needed to prevent wound
\end{enumerate}

\begin{enumerate}
\item Which episodes of season four of game of thrones did michelle maclaren direct
\item Which installment of season four of game of thrones did michelle maclaren direct
\item Which sequence of season four of game of thrones did michelle maclaren direct
\item Which sequence of season quaternity of game of thrones did michelle maclaren direct
\end{enumerate}

\begin{enumerate}
\item She directed oathkeeper and first of his name the fourth and fifth episodes of season four respectively
\item She directed oathkeeper and get-go of his name the fourth and fifth episodes of season four respectively
\item She directed oathkeeper and get-go of his name the fourth and fifth episodes of season four respectively
\item She directed oathkeeper and get-go of his epithet the quartern and one-fifth sequence of season quaternity respectively
\end{enumerate}

\begin{enumerate}
\item Process mining is a family of techniques relating the fields of data science and process management to support the analysis of operational processes based on event logs the goal of process mining is to turn event data into insights and actions process mining is an integral part of data science fueled by the availability of event data and the desire to improve processesprocess mining techniques use event data to show what people machines and organizations are really doing process mining provides novel insights that can be used to identify the executional path taken by operational processes and address their performance and compliance problemsprocess mining starts from event data input for process mining is an event log an event log views a process from a particular angle each event in the log should contain 1 a unique identifier for a particular process instance called case id 2 an activity description of the event that is occurring and 3 a timestamp there may be additional event attributes referring to resources costs etc but these are optional with some effort such data can be extracted from any information system supporting operational processes process mining uses these event data to answer a variety of processrelated questions
\item Summons mining is a family of techniques relating the fields of data science and process management to support the analysis of operational processes based on event logs the goal of process mining is to turn event data into insights and actions process mining is an integral part of data science fueled by the availability of event data and the desire to improve processesprocess mining techniques use event data to show what people machines and organizations are really doing process mining provides novel insights that can be used to identify the executional path taken by operational processes and address their performance and compliance problemsprocess mining starts from event data input for process mining is an event log an event log views a process from a particular angle each event in the log should contain 1 a unique identifier for a particular process instance called case id 2 an activity description of the event that is occurring and 3 a timestamp there may be additional event attributes referring to resources costs etc but these are optional with some effort such data can be extracted from any information system supporting operational processes process mining uses these event data to answer a variety of processrelated questions
\item Summons mining is a family of techniques relating the fields of data science and process management to support the analysis of operational processes based on event logs the goal of process mining is to turn event data into insights and actions process mining is an integral part of data science fueled by the availability of event data and the desire to improve processesprocess mining techniques use event data to show what people machines and organizations are really doing process mining provides novel insights that can be used to identify the executional path taken by operational processes and address their performance and compliance problemsprocess mining starts from event data input for process mining is an event log an event log views a process from a particular angle each event in the log should contain 1 a unique identifier for a particular process instance called case id 2 an activity description of the event that is occurring and 3 a timestamp there may be additional event attributes referring to resources costs etc but these are optional with some effort such data can be extracted from any information system supporting operational processes process mining uses these event data to answer a variety of processrelated questions
\item Summons excavation is a menage of technique relating the plain of information skill and summons direction to living the psychoanalysis of operational process based on effect logs the destination of summons excavation is to round effect information into perceptiveness and activeness summons excavation is an integral region of information skill fueled by the accessibility of effect information and the desire to improve processesprocess excavation technique utilization effect information to appearance what citizenry machine and formation are really doing summons excavation provides novel perceptiveness that can be used to identify the executional route taken by operational process and savoir-faire their operation and conformation problemsprocess excavation get-go from effect information comment for summons excavation is an effect logarithm an effect logarithm aspect a summons from a specific slant each effect in the logarithm should contain single a unique identifier for a specific summons example called suit idaho two an action description of the effect that is occurring and deuce-ace a timestamp there whitethorn be additional effect attribute referring to imagination costs etc but these are optional with some movement such information can be extracted from any data scheme support operational process summons excavation us these effect information to solvent a miscellanea of processrelated dubiousness
\end{enumerate}

\begin{enumerate}
\item Process mining brings together data mining data science and process management to find process efficiencies
\item Summons mining brings together data mining data science and process management to find process efficiencies
\item Summons mining brings together data mining data science and process management to find process efficiencies
\item Summons excavation brings together information excavation information skill and summons direction to discovery summons efficiencies
\end{enumerate}

\begin{enumerate}
\item What are some unique curtain tie backs that you can make yourself
\item What ar some unique curtain tie backs that you can make yourself
\item What are some unique pall tie backs that you can make yourself
\item What are some unique pall link rachis that you can shuffling yourself
\end{enumerate}

\begin{enumerate}
\item There are many items you can use to make a curtain tie back some ideas for this include a chain a circle and pin jute rope wooden beaded rope a necklack or bracelet a door knob a length of a leather belt macrame rope or a string of artificial flowers
\item There ar many items you can use to make a curtain tie back some ideas for this include a chain a circle and pin jute rope wooden beaded rope a necklack or bracelet a door knob a length of a leather belt macrame rope or a string of artificial flowers
\item There are many token you can use to make a curtain tie back some ideas for this include a chain a circle and pin jute rope wooden beaded rope a necklack or bracelet a door knob a length of a leather belt macrame rope or a string of artificial flowers
\item There are many token you can utilization to shuffling a pall link rear some estimation for this include a concatenation a round and tholepin jute rophy wooden beaded rophy a necklack or bangle a threshold node a distance of a leather bash macrame rophy or a chain of artificial flush
\end{enumerate}

\begin{enumerate}
\item What is a dispersive prism
\item What is angstrom dispersive prism
\item In eye a dispersive prism is an optical prism that is used to disperse light that is to separate light into its spectral components the colors of the rainbow different wavelengths colors of light will be deflected by the prism at different angles this is a result of the prism material's index of refraction varying with wavelength dispersion generally longer wavelengths red undergo a smaller deviation than shorter wavelengths blue the dispersion of white light into colors by a prism led sir isaac newton to conclude that white light consisted of a mixture of different colors
\item What is a dispersive prism
\end{enumerate}

\end{document}